\documentclass[a4paper, 12pt]{report}

\usepackage[usenames,dvipsnames]{xcolor}
\usepackage[utf8]{inputenc}				% UTF-8 encoding, so that you can use characters like ç and ã
\usepackage[T1]{fontenc}				% Same, but for output encoding
\usepackage[portuguese,english]{babel}	% Still related to the above
\usepackage{todonotes}
\usepackage{acronym}					% List of acronyms
\usepackage{textcomp} 					% Extra characters
\usepackage{graphicx} 					% \includegraphics{}, the most common command to include images in figures 
\usepackage{titlesec}		 			% To manually format the chapter titles
\usepackage[left=3cm,right=2cm,top=2.5cm,bottom=2.5cm]{geometry} % Margins, as dictated by the rules
\usepackage[nottoc]{tocbibind} 	% Hyperlinks in table of contents, useful for navigation
\usepackage{multirow, makecell, pifont, rotating}
\usepackage[section]{placeins}			% \FloatBarrier, a useful command when your figures are trying to run away
\usepackage{caption}					% For captioning figures
\usepackage{subcaption}					% Subfigures (the subfigure package is deprecated and should not be used)
\usepackage[toc,page]{appendix}			% Appendices
\usepackage{pdfpages}					% Useful when your appendix is a pre-compiled PDF, such as a whole paper
\usepackage{url}						% Useful when one wants to include URLs in the text
\usepackage[
      colorlinks=true,    			%no frame around URL
      urlcolor=black,    			%no colors
      menucolor=black,    			%no colors
      linkcolor=black,    			%no colors
      citecolor=black,    			%no colors
      bookmarks=true,    			%tree-like TOC
      bookmarksopen=true,    		%expanded when starting
      bookmarksnumbered=true, 		%Put section numbers in bookmarks
      hyperfootnotes=true,    		%no referencing of footnotes, does not compile
      pdfpagemode=UseOutlines,    	%show the bookmarks when starting the pdf viewer
      plainpages=false, 			%solve problem ``destination with the same identifier'' warning
      pdfpagelabels				 	%solve problem ``destination with the same identifier'' warning
]{hyperref} 							% So that our citations look good and still work as links
\usepackage{epigraph}					% For your inspirational quote
\usepackage{etoolbox}
\usepackage{float}
\usepackage{arydshln}
\usepackage{tikz,lipsum,lmodern}
\usepackage[most]{tcolorbox}
\usepackage{listings}
\usepackage{lipsum}
\usepackage{amsmath}
\titleformat{\chapter}[hang]
{
  \sffamily
  \Huge
  \bfseries
}
{\thechapter}{0.5em}
{}

\definecolor{codegreen}{rgb}{0,0.3,0}
\definecolor{codegray}{rgb}{0.5,0.5,0.5}
\definecolor{codepurple}{rgb}{0.38,0,0.62}
\definecolor{backcolour}{rgb}{1,1,1}

\newcommand{\MONTH}{%
  \ifcase\the\month
  \or January% 1
  \or February% 2
  \or March% 3
  \or April% 4
  \or May% 5
  \or June% 6
  \or July% 7
  \or August% 8
  \or September% 9
  \or October% 10
  \or November% 11
  \or December% 12
  \fi}
\makeatletter
\newcommand{\YEAR}{\the\year}
\makeatother
\colorlet{shadecolor}{lightgray}

\newcommand{\thesistitle}{Lattice Computation of the Kugo-Ojima Correlation Function} %Your work's title
\newcommand{\statedate}{Coimbra, \MONTH, \YEAR}					% The date, usually "Place, Month Year"
\newcommand{\supervisorname}{Prof Dr. Orlando Olavo Aragão Aleixo e Neves de Oliveira}		% Your supervisor's name
\newcommand{\cosupervisorname}{Dr. Paulo de Jesus Henriques da Silva}	% Your co-supervisor's name, if any.

\begin{document}

\pagenumbering{roman}

% TITLE PAGES
% Uncomment this line when you have your cover ready. An MSWord template is available at that folder.
% You should edit it in MSWord, and then export it into PDF, so we can neatly import it here.
\includepdf[pages={1}]{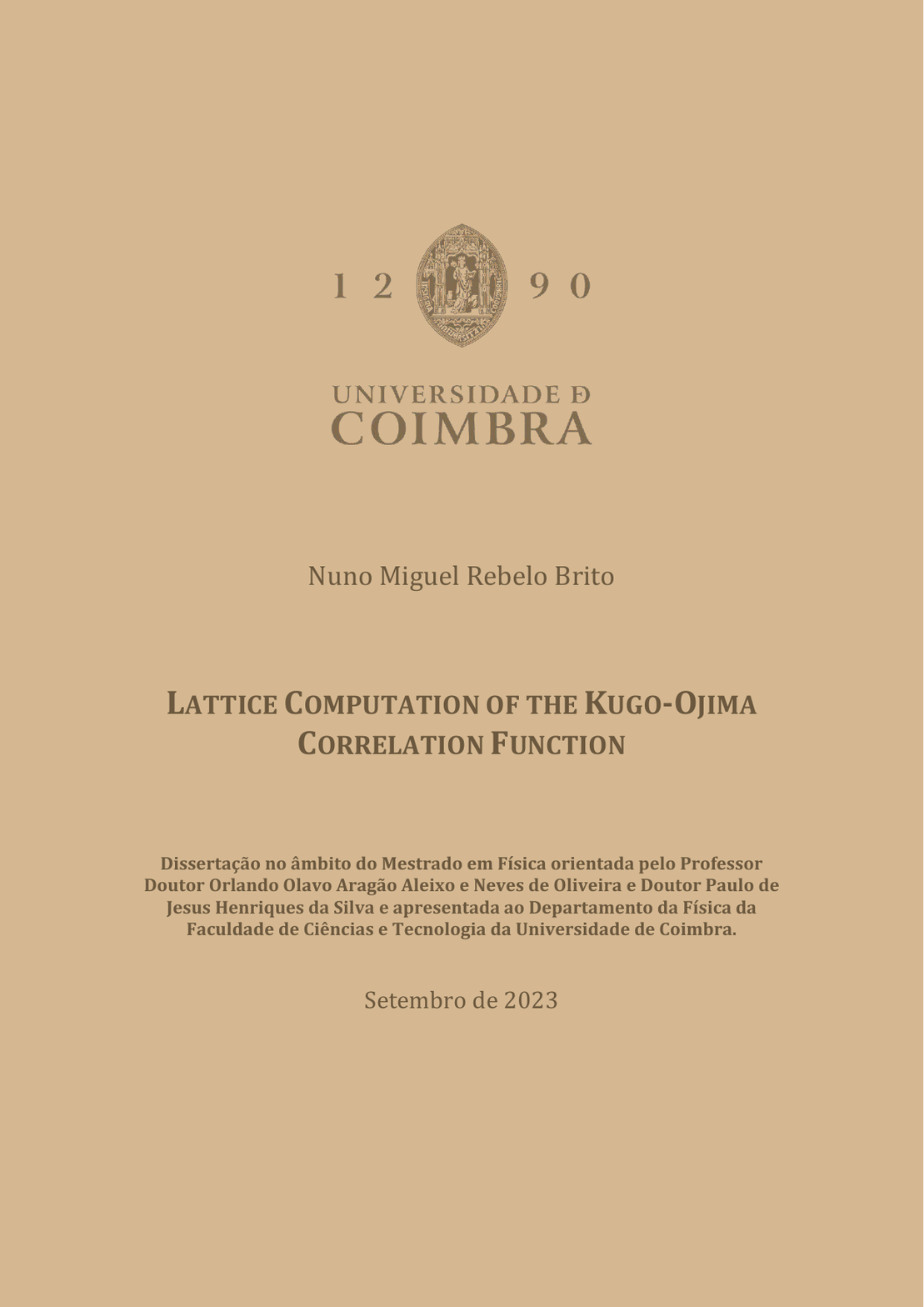}
% Blank page
\newpage
\thispagestyle{empty}
\mbox{}
% Title page
\begin{titlepage}
    \begin{center}
    % UC logo, no name
    \includegraphics[width=0.5\textwidth]{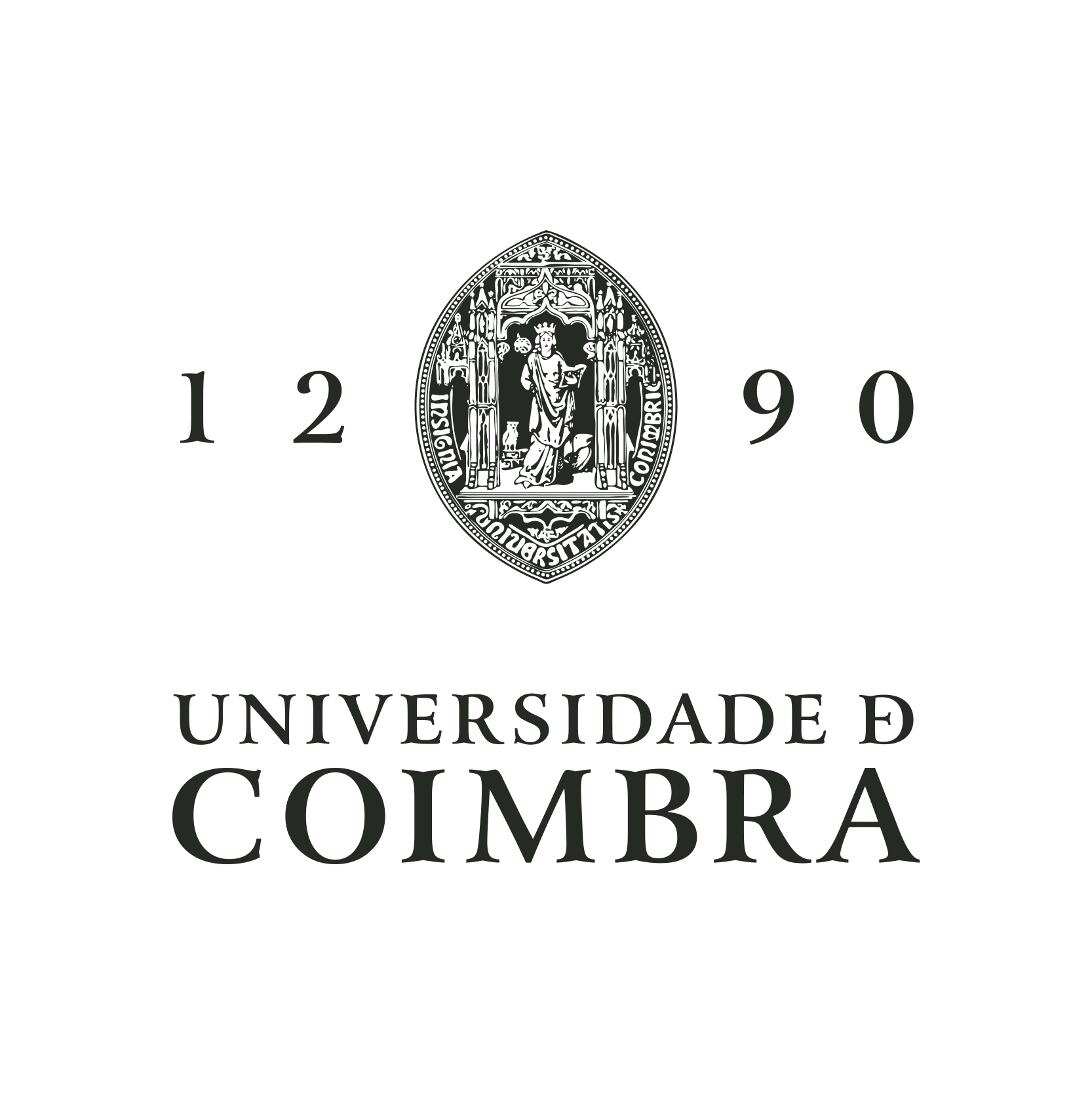}
    
    % Thesis name
    \vspace{0.3cm}
    {\Huge{\textbf{\thesistitle}}\par}
    
    \vspace{1cm}
    {\large{\textbf{Supervisors:}\\\supervisorname\par\cosupervisorname}}
    \vspace{5mm}
    %{\large{\textbf{Supervisor:}\\\cosupervisorname}}
    
    {\large{\textbf{Jury:}
    
    Prof. Dr. Pedro Almeida Vieira Alberto
    
    Dr. Paulo de Jesus Henriques da Silva  
    
    Prof. Dr. David Henk Dudal  
    
    }}
    
    % Final Stuff
    \vfill
    Dissertation submitted in partial fulfillment for the degree of Master of Science in Physics.
    
    \vspace{0.3cm}
    {\large \statedate\par}

    \end{center}
\end{titlepage}

% Blank page
\newpage
\thispagestyle{empty}
\mbox{}

% Acknowledgements
\chapter*{Acknowledgments}
\addcontentsline{toc}{chapter}{Acknowledgements}
I would like to start by expressing my deepest gratitude towards my advisors Prof. Dr. Orlando Oliveira and Dr. Paulo Silva, for their patience and guidance throughout the past year, as well as very valuable advice. I have learned so much about a lot of things, ranging from Physics to the world of Physics research and for that I am very thankful. It is far from easy to work with someone completely new whose work dynamics are unknown and I am aware that especially in my case, it might have been challenging.
Their contributions towards this dissertation were incredibly vital and I dearly appreciate their endless patience.
\par
\noindent
I am also very grateful for the collaboration and fruitful discussions with Dr Joannis Papavassiliou and Dr. Mauricio Ferreira from the University of Valencia.
\par
\noindent
To my friends group \emph{SU(5)} who stuck together throughout the insanity that was the first years of the Master’s Degree, thank you for putting up with me in our darkest hours.
\par
\noindent
To my friends group \emph{Phelps} and \emph{Dia D}, thank you for the good company during this and the last year. It gets quite lonely being a graduate student, we often grow apart from our previous acquaintances.
\par
\noindent
To my closest friend group, thank you, for the good and bad times, for the fun and the boring, for the hype and for the down times. We built an awesome story together and I can only wish that I could have been more present. Special thanks to Benitake, Ramon, Bruno the Giant, Cravo and Afonso and Carrasco.
\par
\noindent
To my colleague Helena, who has been the literal impersonation of "Same" throughout this year. Whether the ship was afloat or sinking we were on it.
\par
\noindent
To T\^{a}nia who has probably been the first person I met who has shown me her unfiltered appreciation for me, something that I had not known how it felt for a long time. Although we might grow apart, you surely left a good mark on me.
\par
\noindent
To all my acquaintances in Coimbra, thank you. I did not hesitate in choosing the Physics Department in Coimbra and I’m glad for that. It was an awesome place to study and to be a student if you know what I mean.
\par
\noindent
Thank you Hakita for designing and making the awesome game that is \emph{ULTRAKILL} which has entertained me so much throughout these last two years.
\par
\noindent
Now, onto the two most important acknowledgements. To my family, who have been undeniably and unconditionally a huge support pillar for me. Thank you Mother for being, objectively might I add, the best mother in the world. Thank you Father for the calm and wise advice, which has helped me become a version of myself that I am very proud of. Thank you Brother for being my life advisor and for the valuable life advice.
\par
\noindent
And finally, to my girlfriend Cláudia, who has undoubtedly changed my life for the good, for the endless support, love and company. This year was very hard, and that difficulty would only increase exponentially if it were not for you. You are most certainly the best thing that has happened to me 
during my time in Coimbra.
\par
\noindent
This work was supported with funds from the Portuguese National Budget through Fundação para a Ciência e Tecnologia (FCT) under the projects UIDB/04564/2020 and UIDP/04564/2020. The author acknowledges the Laboratory for Advanced Computing at the University of Coimbra (http://www.uc.pt/lca) for providing access to the HPC resource \emph{Navigator} that have contributed to the research within this paper. Access to \emph{Navigator} was partly supported by the FCT Advanced Computing Projects 2021.09759.CPCA and 2022.15892.CPCA.A2. The author also acknowledges the computing resources provided by the Partnership for Advanced Computing in Europe (PRACE) initiative under DECI-9 project COIMBRALATT and
DECI-12 project COIMBRALATT2.
\begin{flushright}
    \textit{Torro(z/s)elo, Setembro de 2023}
\end{flushright}

% You can add blank pages here, if you like

% ABSTRACT
\chapter*{Resumo}
\addcontentsline{toc}{chapter}{Resumo}
Atualmente, o confinamento da cor na Cromodinâmica Quântica permanece um mistério do ponto de vista teórico. Até hoje, não foi encontrada prova analítica para o confinamento de cor, e o mecanismo responsável pelo confinamento dos estados com cor do subespaço de estados físicos é ainda desconhecido. Taichiro Kugo e Izumi Ojima propuseram um mecanismo de confinamento baseado na simetria BRST e derivaram assim os requerimentos para a realização deste mecanismo. Um desses
requerimentos, que por sua vez é o menos óbvio, é que uma função de correlação especial, chamada \emph{Função de correlação de Kugo-Ojima} $u(p^2)$ tende para $-1$ na origem $(p^2=0)$.
Esta função pode ser obtida na rede usando a formulação de teorias de gauge na rede.
\par
\noindent
Este trabalho consiste nos resultados da rede para esta função de correlação na gauge de Landau. Apresentam-se os resultados para 4 redes simétricas de grande volume $(32^{4}, 48^{4}, 64^{4}, 80^{4})$ com $\beta=6.0$ na gauge de Landau. Testes relativamente à transversalidade da função de Kugo-Ojima também são feitos assim como também considerações estatísticas dos resultados.
Os resultados apresentam mais provas que o mecanismo de confinamento de Kugo-Ojima não é realizado e que a função de correlação de Kugo-Ojima, na gauge de Landau, é de facto transversa.
Os nossos resultados alinham-se qualitativamente com literatura existente, contribuindo para o nosso entendimento de Cromodin\^{a}mica Qu\^{a}ntica. Esta esforço atual para descrever o fenómeno de confinamento de cor permanece com um desafio central em Física de Partículas.
\par

\textbf{Palavras-Chave:} Confinamento de cor, QCD na rede, Função de Kugo-Ojima, Teoria Quântica de Campos, Gauge de Landau.

% And here

\chapter*{Abstract}
\addcontentsline{toc}{chapter}{Abstract}
As of today, color confinement in Quantum Chromodynamics remains a mystery from the theoretical point of view. So far, no analytical proof of color confinement has been found and the mechanism that confines colored states from the space of physical states is still unknown. Taichiro Kugo and Izumi Ojima proposed such confinement mechanism, using as basis the BRST-symmetry and derived the requirements for the realization of this mechanism. One such requirement, which happens to be the non-trivial one, is that a special correlation
function, the \emph{Kugo-Ojima correlation function} $u(p^2)$, approaches $-1$ at the origin ($p^2=0)$. This correlation function can be obtained on the lattice within the lattice formulation of gauge theories.
\par
\noindent
The present work consists on lattice results for this correlation function on the Landau gauge. We present results obtained from 4 symmetric large volume lattices $(32^{4}, 48^{4}, 64^{4}, 80^{4})$ with $\beta = 6.0$ on the Landau gauge. A test on the transversality of the Kugo-Ojima correlation function is also performed, along with some statistical considerations of the results.
The results present further evidence that the Kugo-Ojima confinement scenario is not realized on the lattice and that the Kugo-Ojima correlation function, in the Landau gauge, is transverse.
Our findings align qualitatively with existing literature, contributing to our understanding of Quantum Chromodynamics. This ongoing pursuit to unravel color confinement remains a central challenge in particle physics.
\par

\textbf{Keywords:} Color Confinement, Lattice QCD, Kugo-Ojima function, Quantum Field Theory, Landau Gauge.

% And here as well
\newpage\null\thispagestyle{empty}\newpage

% INSPIRATIONAL QUOTE
% Setup
\setlength\epigraphwidth{12cm}
\setlength\epigraphrule{0pt}
\makeatletter
\patchcmd{\epigraph}{\@epitext{#1}}{\itshape\@epitext{#1}}{}{}
\makeatother
% Actual Quote
\vspace*{\fill}
\epigraph{Luck is what happens when preparation meets opportunity.}
{Lucillius Seneca}
\vspace*{\fill}
\newpage\null\thispagestyle{empty}\newpage
% TABLE OF CONTENTS
\pagestyle{plain}
\tableofcontents
% LIST OF ACRONYMS
%\chapter*{List of Acronyms}
%\addcontentsline{toc}{chapter}{List of Acronyms}
%\input{list_acronyms}
% LIST OF FIGURES
\listoffigures
% LIST OF TABLES
\listoftables
% BODY
\newpage
\thispagestyle{empty}
\mbox{}
\pagenumbering{arabic}	% Arabic numbering starts

% For each chapter, you should have a bit of code that looks like this:
% \label allows you to later \ref that chapter.
% \input includes a different .tex file, so that you can have you dissertation
% neatly partitioned into several files. I recommend one file per chapter.

%-----------------CHAPTERS--------------------

%-----CHAPTER1-INTRODUCTION-----
\chapter{Introduction}
\label{chap:introduction}
Particle Physics is the area of physics concerned with describing the dynamics of matter at the atomic and sub-atomic level. One of its most successful theories is the Standard Model, which provides a description for three of the four fundamental interactions in nature: strong, weak and electromagnetic interaction. The fourth fundamental interaction, gravity, still lacks a quantum description.
\par
\noindent
The Standard Model is a gauge theory based on the gauge group $SU(3) \otimes SU(2) \otimes U(1)$ and it provides an unified description of the dynamics and symmetries of fundamental particles. A fundamental symmetry of the Standard Model Lagrangian is the gauge symmetry that is associated with the gauge group. Indeed, the equations of motion derived from the Lagrangian are invariant under gauge transformations.  
\par
\noindent
The Standard Model Lagrangian is divided into sectors, each of which is responsible for the description of different fundamental particles and their interactions. The electromagnetic and the weak interaction are unified into the electroweak sector, with gauge group $SU(2) \otimes U(1)$.  The strong sector, associated with the gauge group $SU(3)$, describes the interactions between quarks and gluons. The theory of the strong interaction is called Quantum Chromodynamics (QCD). 
\par
\noindent
Quantum Electrodynamics (QED) is the theory of the electromagnetic interaction and it defines the interactions of charged particles such as electrons. The perturbative description of QED has provided highly satisfactory experimental predictions, but the same case cannot be said for QCD. Many features of QCD can only be described in a non-perturbative scheme, one such feature is color confinement. Color is a property inherent to the elemental particles of QCD that can be thought of as the analogue of the electric charge of QED. Color confinement consists of how the fundamental quanta of QCD are not observed asymptotically unlike the photon or the electron. More specifically, colored particles are not observed experimentally as free particles. 
\par
\noindent
This leads to the hypothesis that colored states do not belong to the physical Hilbert space of QCD, similar to how longitudinally polarized photons are excluded from the physical subspace in the Gupta-Bleuler quantization of QED \cite{Gupta1950, Bleuler1950}. However, as of today, a theoretical proof of color confinement does not exist but there have been proposals for a color confinement mechanism. One such proposition was provided in 1979 by Taichiro Kugo and Izumi Ojima \cite{KugoOjima79}. 
\par
\noindent
Their work extended well beyond adressing color confinement; they provided a thorough and detailed description of the Canonical Quantization of Yang-Mills theories. The color confinement mechanism proposed by them used the BRST-symmetry \cite{BECCHI1976287,tyutin2008gauge}, a generalized type of gauge symmetry, to define the Hilbert space of physical states. Then, they prove that every state in this space is a color singlet, but only if two requirements are satisfied. One of these requirements
is that the color symmetry remains unbroken, which they assume given the observed phenomenon of color confinement. The other is that a certain correlation function, called the \emph{Kugo-Ojima correlation function} $u(p^2)$ converges to $-1$ at the origin. Kugo also showed that if this requirement is satisfied, the ghost dressing function would also diverge \cite{Kugo1995} at the origin. 
\par
\noindent
The interest in this function spans beyond color confinement. The function is conceptually connected with the Gribov-Zwanziger framework \cite{zwanziger1989local,Dudal_2009,Kondo_2009} and it is a necessary ingredient for other formulations of Yang-Mills theories (see \cite{Aguilar_2009} or Chapter 2 of \cite{Ferreira_2023} and references therein).
\par
\noindent
To study this function, non-perturbative methods must be used, such as the Lattice formulation of Gauge theories and the Dyson-Schwinger equations. The Lattice formulation discretizes space-time, providing a very convenient framework to simulate Gauge theories on a computer and will be the one we will use. The computation of this function, as well as its properties, is the focus of this work. We intend to provide, using the Lattice formulation of gauge theories, a numerical computation of the
Kugo-Ojima correlation function using a large lattice volume.
\par
\noindent
This work is organized as follows:
\noindent In the next chapter, we give a brief introduction of Quantum Chromodynamics. In the third chapter, the background of the Kugo-Ojima correlation function is explained and the Kugo-Ojima confinement scenario's necessary ingredients are stated. In the fourth chapter, we present the lattice formulation of Gauge Theories and use this formulation to compute the Kugo-Ojima function on the Lattice. Finally, we give a short insight into the obtained results as well as plans for the future.

% Add your necessary chapters

\chapter{Basics of Quantum Chromodynamics}
\label{chap:basics_qcd}
In this chapter, the basic definitions of QCD are given, namely the Lagrangian and its symmetries. The path-integral formulation of field theories is also discussed, as well as how to extract the Green's functions of the theory. The procedures of renormalization and regularization are briefly discussed. 
\section{QCD Lagrangian} 
\label{sec:QCD_Lagrangian}
Quantum Chromodynamics is formulated under the Lagrangian framework. This framework allows the description of the dynamics of the theory as well as the underlying symmetries. At its core is the Lagrangian density:
\begin{equation}
    \label{eq:qcd_lagrangian}
    \mathcal{L}_{QCD} = \sum _{f} \bar{\psi}_{f}(i\gamma^{\mu}\nabla_{\mu} - m_{f} )\psi_{f} -\frac{1}{4}F^{a}_{\mu \nu}F^{a \mu \nu }. 
\end{equation}
\noindent
Here, $\psi_{f}$ denotes the flavored f-quark spinor field, $\bar{\psi}_{f} = \psi_{f}^{\dagger}\gamma^{0}$.
We use latin indices to denote the color component of the fields, and greek ones to denote the Lorentz component. Additionally, if any indices (greek or latin) are repeated, a sum over its values is to be intended. 
The covariant derivative $\nabla_{\mu}$ is given by:
\begin{equation}
    \nabla_{\mu}\psi \equiv \nabla_{\mu}^{ab} \psi^{b} = (\delta^{ab}\partial_{\mu}+ ig[t^{c}]^{ab}A^{c}_{\mu})\psi^{b}.
\end{equation}
The real-valued fields $A^{c}_{\mu}$ are the components of the linear combination of the gauge field $A_{\mu}(x) \equiv A_{\mu}^{c}(x) t^{c}$. This gauge field is an element of the SU(3) Lie Algebra, and its introduction is necessary to maintain the gauge invariance of the theory. The $t^{c}$ are the 8 generators of a given representation of Lie Algebra of SU(3) ($c = 1, \dots, 8)$. This algebra is characterized by the following relations:
\begin{equation}
    \label{eq:lie_algebra_su3}
    [t^{a},t^{b}] = i f^{abc}t^{c},  \qquad \text{Tr}(t^{a} t^{b}) = R\delta^{ab}. 
\end{equation}
\noindent
In this work, the fundamental and the adjoint representations will be considered.
The value of $R$ also depends of the representation (for SU(N), $R = \frac{1}{2}$ for the fundamental representation, $R=N$ for the adjoint representation).  So, we set the following notation for the covariant derivative in these two representations:
\begin{equation}
    \begin{aligned}
    \label{eq:covariant_derivatives}
        [\nabla_{\mu}(x)]^{ab} \equiv \left(\delta^{ab} \frac{\partial}{\partial x^{\mu}}+ ig \frac{[\lambda^{c}]^{ab}}{2}A^{c}_{\mu}(x)\right), \\
        [D_{\mu}(x)]^{ab} \equiv \left(\delta^{ab}\frac{\partial}{\partial x^{\mu}}+ g f^{abc}A^{c}_{\mu}(x)\right).
    \end{aligned}
\end{equation}
\noindent
We reserve the notation $t^{a}$ for the fundamental representation while for the adjoint representation we substitute explicitly $[t^{a}]^{bc} = -i f^{abc}$. In the case of the Lagrangian density \eqref{eq:qcd_lagrangian}, the covariant derivative of the spinor quark field is in the fundamental representation. The Field Strength tensor $F^{a}_{\mu \nu}$ is given by:
\begin{equation}
    \label{eq:field_strength_tensor}
    F^{a}_{\mu \nu} = \partial_{\mu} A^{a}_{\nu} - \partial_{\nu} A^{a}_{\mu} + g(A_{\mu} \times A_{\nu})^{a} \quad \text{ where } \quad (A_{\mu} \times A_{\nu})^{a} \equiv
f^{abc}A^{b}_{\mu}A^{c}_{\nu}.
\end{equation}
\noindent
Equipped with the Lagrangian Density, the action is written as:
\begin{equation}
    S[\psi, \bar{\psi}, A_{\mu}] = \int d^{4}x \text{  } \mathcal{L}[\psi, \bar{\psi}, A_{\mu}]
\end{equation}
and we obtain the respective equations of motion for the fields setting the first variation of the action to zero $\delta S = 0$:
\begin{equation}
    \begin{aligned}
        \label{eq:QCD_equations_of_motion}
        \left(D_{\nu} F^{\mu \nu}\right)^{a} = \bar{\psi} \gamma^{\mu}t^{a}\psi, \\
        \left(i\gamma^{\mu}\nabla_{\mu} + m\right)^{ab}\psi^{b} = 0, \\
        \bar{\psi}^{a}\left(i \gamma^{\mu}\nabla_{\mu} + m\right)^{ab} = 0.
    \end{aligned}
\end{equation}
\noindent
The Lagrangian density also reflects the underlying symmetries of the theory. QCD is a SU(3) Non-Abelian gauge theory, meaning that \eqref{eq:qcd_lagrangian} is invariant under local SU(3) transformations:
\begin{equation}
 \begin{aligned}
    \label{eq:gauge_transform}
    \psi(x) &\longrightarrow \psi^{G}(x) = G(x) \psi(x) \nonumber, \\
    \bar{\psi}(x) &\longrightarrow \bar{\psi}^{G}(x) = \bar{\psi}(x) G^{\dagger}(x), \\ 
    A_{\mu} (x) &\longrightarrow A_{\mu}^{G}(x) = G(x)A_{\mu}(x)G^{\dagger}(x) - \frac{i}{g} G(x)\partial_{\mu}G^{\dagger}(x). \nonumber  
\end{aligned}  
\end{equation}
\noindent
Where $G(x) = e^{ig \alpha^{a}(x)t^{a}}$ is a element of the SU(3) Lie Group, so that $G^{-1}(x) = G^{\dagger}(x)$. It is often simpler to workout the derivation for a infinitesimal gauge transformation since we can build the finite transformation by sucessively applying infinitesimal transformations. The infinitesimal version of the transformation laws in \eqref{eq:gauge_transform} become:
\begin{equation}
    \label{eq:gauge_transform_infinitesimal}
    G(x) \approx (\boldsymbol{1} + ig\alpha^{a}(x)t^{a} + \mathcal{O}(g^2)) \implies 
    \begin{cases}
        & \psi^{G}(x) = (\boldsymbol{1} + ig\alpha^{a}(x)t^{a} + \mathcal{O}(g^2))\psi(x), \\ 
        & \bar\psi^{G}(x) = \bar{\psi}(x)(\boldsymbol{1} - ig\alpha^{a}(x)t^{a} + \mathcal{O}(g^2)), \\
        & \left(A^{G}\right)^{a}_{\mu}(x) = A^{a}_{\mu}(x) + \frac{1}{g} \partial_{\mu} \alpha^{a}(x) + f^{abc}A^{b}_{\mu}(x)\alpha^{c}(x).
    \end{cases}
\end{equation}
The components $\left(A^{G}\right)^{a}_{\mu}(x)$ are calculated by projecting $A_{\mu}$ onto the generators $t^{a}$ using the trace property in \eqref{eq:lie_algebra_su3} in the following way:
\begin{equation}
    \label{eq:components_of_lie_algebra}
    \left(A^{G}\right)^{a}_{\mu}(x) = \frac{1}{R} \operatorname{Tr}\{A^{G}_{\mu}(x) t^{a}\}.
\end{equation}

\section{Path Integral Formulation}%
\label{sec:Path Integral Formulation}

The Path Integral formulation of quantum mechanics was originally created by Feynman \cite{feynman2010quantum}. This formulation allows the generalization of the principle of least action from classical mechanics into quantum mechanics, in a way such that every possible trajectory contributes (or interferes) to the probability amplitude of a given process. In the path integral formulation of quantum field theory, the central object is the \emph{generating functional} \cite{Greiner1996}, which is the vacuum-vacuum amplitude:
\begin{equation}
    \label{eq:generic_generating_functional}
    \langle 0 \mid 0 \rangle_{J} = Z[J_{1},\cdots, J_{n}] = \int \mathcal{D}\phi_{1} \dots \mathcal{D}\phi_{n} \text{ } e^{iS[\phi_{1}, \dots ,\phi_{n}] + i \int d^{4}x \sum_{i} \phi_{i}(x) J_{i}(x)}
\end{equation}
where $J_{i}$ is the source of the field $\phi_{i}$. The \emph{Green's functions} (or correlation functions) of the theory can be derived through consecutive functional derivation of \eqref{eq:generic_generating_functional}:
\begin{equation}
\begin{aligned}
    \label{eq:green_functions}
     \langle 0 | T(\phi_{1}(x_{1}) \cdots \phi_{n}(x_{n})) |0 \rangle &= \frac{\delta^{n}Z[J_{1},\cdots, J_{n}]}{\delta J_{1}(x_{1}) \cdots \delta J_{n}(x_{n})}\Bigg|_{J=0} \equiv G^{(n)}(x_{1},\dots,x_{n}) \nonumber \\ &=\frac{1}{Z[0,\dots,0]} \int\mathcal{D}\phi_{1} \dots \mathcal{D}\phi_{n} \text{ }(\phi_{1}(x_{1}) \cdots \phi_{n}(x_{n})) \text{  }e^{iS[\phi_{1},\cdots, \phi_{n}]}
\end{aligned}
\end{equation}
\noindent
where $T$ is the time-ordering operator. These functions are objects of interest since their knowledge allows us to extract relevant properties and quantities of the system \cite{fetter2012quantum}. As an example, Feynman diagrams are built using these functions and allow writing diagramatically probability amplitudes of relevant processes of the theory. It is also possible to define the Green's functions in momentum space:
\begin{equation}
    \begin{aligned}
    \label{eq:green_functions_momentum_space}
    G^{(n)}(p_{1},\dots,p_{n})\left(2\pi\right)^{4} \delta(p_{1}+\dots+p_{n}) = \\ 
        = \int d^4{x}_{1}\dots d^4{x}_{n} \text{ } e^{-i(p_{1}x_{1}+\dots+ p_{2}x_{2})}G^{(n)}(x_{1},\dots,x_{n}). 
    \end{aligned}
\end{equation}
\noindent
The momentum space Green's functions are often preferred instead of \eqref{eq:green_functions}, since their expressions are often simpler than their coordinate-space counterpart. As a result, the expressions for Feynman diagrams are further simplified \cite{fetter2012quantum}. The delta function included on the left-hand side accounts for momentum conservation.
\subsection{Euclidean Field Theory}%
\label{sub:Euclidean Field Theory}
Quantities such as \eqref{eq:generic_generating_functional} cannot be evaluated exactly most of the times. The integrand contains an exponential with an oscillatory phase which rises convergence problems in \eqref{eq:generic_generating_functional}\footnote{There are some exceptions to this, such as Free Field theories where we can use some the functional version of some gaussian identities.}. Consequently, it is often performed a \emph{Wick's rotation}, which consists of an analytical continuation into the complex plane, by setting:
\begin{equation}
    \label{eq:complex_times}
    t = e^{-i\frac{\pi}{2}}t_{E} = -i t_{E}
\end{equation}
which implies that the generating functional \eqref{eq:generic_generating_functional} is now written as:
\begin{equation}
    \label{eq:euclidean_generating_functional}
    Z[J_{1},\cdots, J_{n}] = \int \mathcal{D}\phi_{1} \dots \mathcal{D}\phi_{n} \text{ } e^{-S[\phi_{1}, \dots ,\phi_{n}] + \int d^{4}x \sum_{i} \phi_{i}(x) J_{i}(x)}.
\end{equation}
\noindent
If the Euclidean version of the theory satisfies the Osterwalder-Schrader's axioms \cite{osterwalder1975axioms}, then the corresponding Green's functions of the theory may be analytically continued back into the Minkowskian space-time theory. So, one can work with the theory in its Euclidean version \eqref{eq:euclidean_generating_functional}, and then return to
the Minkowski space-time. However, the analytical continuation back to the Minkowski space-time might not be valid in the non-perturbative regime, but if we work with the Euclidean version of the theory, we need not to return to the Minkowski space-time.
\subsection{Faddeev-Popov Procedure}%
\label{sub:Faddeevv-Popov Procedure}
As a consequence of the gauge invariance, the action is also gauge invariant:
\begin{equation}
    \label{eq:Action Invariance}
    S[A_{\mu}^{G},\psi^{G},\bar{\psi}^{G}] = S[A_{\mu},\psi,\bar{\psi}] 
\end{equation}
where $A_{\mu}^{G}$ is the transformed gauge field as in \eqref{eq:gauge_transform}. For simplicity, let's consider only the gauge part of the generating functional:
\begin{equation}
    \label{eq:Example Generating Functional}
    Z_{G}[\omega^{\mu}] = \int \Pi_{\mu}\mathcal{D}A_{\mu} e^{-S[A_{\mu}] - \int d^{4}x \omega^{\mu}(x) A_{\mu}(x)}.
\end{equation}
\noindent
The problem with formula \eqref{eq:Example Generating Functional} is that for a given $A_{\mu}$ there are an infinite number of gauge-related gauge fields, which spoils the functional integration $\Pi_{\mu}\mathcal{D}A_{\mu}$. To proceed, one has to restrict the integration so that we have a single representative per gauge orbit\footnote{A gauge orbit is the set of gauge fields related by a gauge transformation.}. Faddeev and Popov \cite{FaddeevPopov1967} have shown how to achieve such by imposing a condition on the space of the gauge fields $A_{\mu}$ of the type:
\begin{equation}
    \label{eq:gauge_fix_condition}
    F[A;x] = 0.
\end{equation}
In this work, the Landau Gauge is considered, which means imposing: 
\begin{equation}
    \label{eq:landau_gauge}
    F[A;x] = \partial^{\mu} A_{\mu}(x) \implies F[A;x]^{a} \equiv \partial^{\mu} A_{\mu}^{a}(x) = 0. 
\end{equation}
The details regarding this procedure may be found in the original paper by Faddeev and Popov \cite{FaddeevPopov1967} or in any standard textbook \cite{Greiner1996,Ryder1985,peskin2018}. The resulting generating functional after the Faddeev-Popov procedure is written as:
\begin{equation}
 \begin{aligned}
    \label{eq:qcd_generating_functional}
    Z[\eta,\bar{\eta},\omega^{\mu}, \sigma, \bar{\sigma}] = \int \mathcal{D}A_{\mu} \mathcal{D}\psi \mathcal{D} \bar{\psi} \mathcal{D} c \mathcal{D} \bar{c} \operatorname{exp}\{i S[\psi, \bar{\psi}, A_{\mu}]\} \\ 
    \operatorname{exp}\left\{ i \int d^{4}x \left(\mathcal{L}_{FP} + \mathcal{L}_{GF} + \omega^{a \mu} A_{\mu}^{a} + \bar{\eta}^{a}\psi^{a} + \bar{\psi}^{a}\eta^{a} + \bar{c}^{a}\sigma^{a}+ \bar{\sigma}^{a} c^{a} \right)\right\} 
\end{aligned}
\end{equation}
where:
\begin{equation}
\begin{aligned}
    \label{eq:fp_gf_lagrangian_terms}
    \mathcal{L}_{FP} = \bar{c}^{a} \partial^{\mu} \left(D_{\mu} c\right)^{a}, \\
    \mathcal{L}_{GF} = -\frac{1}{2\xi} (\partial^{\mu} A_{\mu}^{a})^2.
\end{aligned}
\end{equation}
The fields $c$ and $ \bar{c}$ are called \emph{ghost fields} and they are Grassmann variables. The inclusion of these fields allows writing the restriction that \eqref{eq:gauge_fix_condition} imposes on the integration $\Pi_{\mu} \mathcal{D}A_{\mu}$ as a simple exponential factor in the generating functional \eqref{eq:qcd_generating_functional} \cite{peskin2018}.
The \emph{effective Lagrangian Density} is defined as:
\begin{equation}
    \label{eq:effective_lagrangian}
    \mathcal{L}_{\text{eff}} = \mathcal{L}_{QCD} + \mathcal{L}_{FP} + \mathcal{L}_{GF}.
\end{equation}
\subsection{The Gribov Copies Problem}%
\label{sub:The Gribov Copies Problem}
However, it has been proven that for the case of non-perturbative QCD, as well as with other Non-Abelian Gauge Theories, that the Faddeev-Popov procedure fails. Namely, \eqref{eq:landau_gauge} is not able to select a single representative per orbit, which is also known as the Gribov Copies Problem \cite{Gribov1978}. Furthermore, it has been proven that for a four-dimensional sphere \cite{Singer1978} and for a four-dimensional torus\footnote{which would correspond to our use case, an Euclidian theory with periodic boundary conditions.} \cite{KillingBack1984}, it is impossible to find a local gauge-fixing condition that selects a single representative per orbit. The condition \eqref{eq:landau_gauge} alone defines the transverse hyperplane of configurations:
\begin{equation}
    \label{eq:transverse_hyperplane}
    \Gamma \equiv \{A: \partial^{\mu} A_{\mu} = 0\}.
\end{equation}
Gribov suggested modifying the condition \eqref{eq:landau_gauge}  further restricting this region to the \emph{Gribov's Region}, using the Faddeev-Popov operator $M[A;x,y]$ \cite{Gribov1978}:
\begin{equation}
    \label{eq:gribov_region}
    \Omega = \{A \in \Gamma: M[A]^{ab} \geq 0\}, \quad M[A;x,y]^{ab} = \frac{\delta F[A^{G};x]^{a}}{\delta G^{b}(y)} = -\partial^{\mu}D_{\mu}^{ab}[A]. 
\end{equation}
However, this region is still not exempt from Gribov Copies \cite{zwanziger1989local,zwanziger1993renormalizability}, implying that we need to provide an additional condition. This motivates the definition of the \emph{fundamental modular region} $\Lambda$:
\begin{equation}
    \label{eq:fundamental_modular_region}
    \Lambda = \{A \in \Omega: A = \operatorname{min} F_{A}[G] \}, \quad F_{A}[G] = \int d^{4}x \sum_{\mu}\operatorname{Tr}\left\{A^{G}_{\mu}(x) A^{G}_{\mu}(x)\right\}
\end{equation}
where only absolute minima are considered. The fundamental modular region $\Lambda$ is free of Gribov Copies. Each gauge orbit has a single representative in the interior %\footnote{Unlike the boundary $\partial \Lambda$, which has degenerate absolute minima that constitute Gribov copies.\cite{VANBAAL1992259,vanbaal1995global,van1994zooming}.} 
of $\Lambda$ \cite{VANBAAL1992259,dell1991every}. This choice of gauge is often called \emph{Minimal Landau Gauge}. It should be noted that the Gribov's region contains all the minima
(global or local) of $F_{A}[G]$ whereas $\Lambda$ only contains the absolute minima. 
\section{Regularization and Renormalization}%
\label{sec:Regularization and Renormalization}
When computing the Green's functions \eqref{eq:green_functions}, one often ends with a divergent expression. Loop diagrams contain integrations in momentum space, which for most theories diverge (ultraviolet divergences). To circumvent this issue, a \emph{regularization} procedure followed by \emph{renormalization} is performed.  
\subsection{Regularization}%
\label{sub:Regularization}
Regularization consists in isolating the divergences of the theory using a regulator. This allows for the identification of the type of divergence of the theory. At the perturbative regime, many regularization schemes may be used such as Dimensional Regularization or the Pauli-Villars Regularization. For the non-perturbative regime, a widely used regularization scheme is the Lattice Regularization, which is the scheme in this work. This scheme is covered in
chapter 3.
\subsection{Renormalization}%
\label{sub:Renormalization}
Once the theory is regularized, renormalization may take place by adding \emph{counter-terms} to the Lagrangian density that effectively cancel our these divergences, while keeping the overall form of the Lagrangian unchanged which allows for the conservation of the dynamics of the theory. Since the overall form of the Lagrangian is preserved, the renormalization procedure is equivalent to a multiplicative rescaling of the fields. A possible effective Lagrangian is given by \cite{Alkofer_2001}:
\begin{equation}
\begin{aligned}
    \mathcal{L}_{\mathrm{eff}}^r= & Z_3 \frac{1}{2} A_\mu^a\left(-\partial^2 \delta_{\mu \nu}-\left(\frac{1}{Z_3 \xi_r}-1\right) \partial_\mu \partial_\nu\right) A_\nu^a \nonumber \\ 
    & +\widetilde{Z}_3 \bar{c}^a \partial^2 c^a+\widetilde{Z}_1 g_r f^{a b c} \bar{c}^a \partial_\mu\left(A_\mu^c c^b\right)-Z_1 g_r f^{a b c}\left(\partial_\mu A_\nu^a\right) A_\mu^b A_\nu^c \nonumber \\ 
    & +Z_4 \frac{1}{4} g_r^2 f^{a b e} f^{c d e} A_\mu^a A_\nu^b A_\mu^c A_\nu^d+Z_2 \bar{\psi}\left(-\gamma_\mu \partial_\mu+Z_m m_r\right) \psi \nonumber \\ 
    & -Z_{1 F} i g_r \bar{\psi} \gamma_\mu T^a \psi A_\mu^a .
\end{aligned}
\end{equation}
Where the multiplicative factors are related by:
\begin{equation}
    \label{eq:z_factor_relations}
    Z_{g}g_{r} = g_{0}\quad Z_{1 F}=Z_g Z_2 Z_3^{\frac{1}{2}}, \quad Z_1=Z_g Z_3^{\frac{3}{2}}, \quad \tilde{Z}_1=Z_g \tilde{Z}_3 Z_3^{\frac{1}{2}}, \quad Z_4=Z_g^2 Z_3^2.
\end{equation}
These relations can be derived from the Slavnov-Taylor identities \cite{Alkofer_2001}.
The renormalized Green's functions are equal to the unrenormalized ones with a multiplicative Z factor:
\begin{equation}
    \label{eq:Renormalized Green's Functions}
    \langle 0 | T(\phi_{1}(x_{1}) \cdots \phi_{n}(x_{n})) |0 \rangle_{R} = Z \cdot \langle 0 | T(\phi_{1}(x_{1}) \cdots \phi_{n}(x_{n})) |0 \rangle.
\end{equation}
The determination of the $Z$ factors requires a choice of a renormalization scheme. Among the most used examples are the Minimal Subtraction scheme (MS) and the MOMentum space subtraction (MOM). The former defines the $Z$ factors in such a way that the counter-terms introduced cancel out only the pole part of the respective propagators and vertices whereas the former sets the renormalized 2-point and 3-point Green's functions to be equal to its tree-level form at a given momentum value
$\mu^2$ \cite{celmaster1979}. An example of this is the determination of $Z_{3}$ for the gluon propagator:
\begin{equation}
    \label{eq:mom_scheme_gluon}
    D_{R}(\mu^2,\mu^2) = Z_{3}(\mu^2)D(\mu^2) = \frac{1}{\mu^2}, \quad D_{\mu \nu}^{ab}(p^{2}) = \delta^{ab} D(p^2)\left( g_{\mu \nu} + (\xi - 1 ) \frac{p^{\mu} p^{\nu}}{p^2}\right)
\end{equation}
where $D^{ab}_{\mu \nu}(p^2)$ is the Green's function for the gluon propagator in momentum space and $D(p^2)$ is the corresponding form factor. Note that an extra dependence on the renormalization point $\mu^2$ was added to the renormalized propagator $D_{R}(p^2,\mu^2)$.

\chapter{Kugo Ojima Confinement Scenario}%
\label{cha:Kugo Ojima Confinement Scenario}
Following the Faddeev-Popov procedure, the effective Lagrangian Density of the theory \eqref{eq:effective_lagrangian} contains the ghost fields. It is known that these fields have the wrong spin-statistics relations \cite{peskin2018}. These fields represent unphysical states, and should not contribute to the physical content of the theory (such as scattering amplitudes). Furthermore, it has not been observed an asymptotic free colored state (e.g. free quarks or gluons). This all suggests that there must
exist a confinement mechanism that prevents these states to contribute to the S-matrix. While formulating the canonical formalism of non-Abelian gauge theories, Taichiro Kugo and Izumi Ojima \cite{KugoOjima79} provided an hypothesis to such mechanism and it is the theme of this chapter. We will present only the key details, the full treatment can be found in \cite{KugoOjima79,Nakanishi}. 
\section{Preliminary Properties}
\label{sec:Preliminary Properies}
In order to discuss the Kugo-Ojima confinement scenario, it is convenient to switch temporarily to the canonical formalism. Additionally, the auxiliary hermitian fields (also called the Nakanishi-Lautrup auxiliary fields\footnote{The auxiliary fields $B^{a}$ originally appeared as a consistent way to perform canonical quantization of electromagnetism in the Landau gauge \cite{Nakanishi66}.}) $B^{a}$ are introduced in the theory by changing the gauge fixing term into:
\begin{equation}
    \label{eq:gauge_fixing_term_canonical}
    \mathcal{L}_{GF} = \frac{\xi}{2} \left(B^{a}\right)^{2} - B^{a}\partial^{\mu}A_{\mu}^{a}.
\end{equation}
The introduction of the auxiliary fields $B^{a}$ do not change the dynamics since no new degrees of freedom are introduced with their addition. It is also important to redefine the anti-ghost as \cite{KugoOjima79}:
\begin{equation}
    \label{eq:ghost_anti_hermitian}
    \bar{c}^{a} \longrightarrow i \bar{c}^{a} \implies (c^{a})^{\dagger}= c^{a}, \quad (\bar{c}^{a})^{\dagger} = \bar{c}^{a}
\end{equation}
which implies that the Faddeev-Popov term is also changed to:
\begin{equation}
    \label{eq:faddeev_popov_term}
    \mathcal{L}_{FP}=i \bar{c}^{a} \partial^{\mu}\left(D_{\mu}c\right)^{a}
\end{equation}
so that the overall Lagrangian has the following form:
\begin{equation}
    \begin{aligned}
    \label{eq:canonical_qcd_lagrangian}
    \mathcal{L}_{\text{eff}} = -\frac{1}{4} (F^{a})^{\mu \nu} (F^{a})_{\mu \nu} + \sum _{f} \bar{\psi}_{f}(i\gamma^{\mu}\nabla_{\mu} - m_{f} )\psi_{f} + \\
        + \frac{\xi}{2} \left(B^{a}\right)^{2} - B^{a}\partial^{\mu}A_{\mu}^{a} + i \bar{c}^{a} \partial^{\mu}\left(D_{\mu}c\right)^{a}.
    \end{aligned}
\end{equation}
The canonical conjugate momenta are:
\begin{equation}
    \label{eq:canonical_conjugate_momenta}
    \begin{aligned}
        & \pi_{B}^{a} = \frac{\partial \mathcal{L}_{\text{eff}}}{\partial \dot{B}^{a}} = -A_{0}^{a}, \quad \pi_{A_{k}}^{a} = \frac{\partial \mathcal{L}_{\text{eff}}}{\partial \dot{A}_{k}^{a}} = F_{0 k}^{a}, \quad \pi_{c}^{a} = \frac{\partial}{\partial \dot{c}^{a}}\mathcal{L}_{\text{eff}} = i\dot{\bar{c}}^{a}. \\
    & \pi_{\bar{c}}^{a} = \frac{\partial}{\partial \dot{\bar{c}}^{a}}\mathcal{L}_{\text{eff}} = -i(\dot{c} + g f^{abc}A_{0}^{b}c^{c}), \quad \pi_{\psi} = \frac{\partial }{\partial \dot{\psi}}\mathcal{L}_{\text{eff}} = \bar{\psi}(i \gamma^{0}).
    \end{aligned}  
\end{equation}
Note that for Grassmann variables, we adopt the left derivative definition. Having the canonical conjugate momenta defined, we can establish the usual (equal time) commutation relations:
\begin{equation}
    \label{eq:commutation_relations}
    \begin{aligned}
        & [\pi^{I}_{\Phi}(x), \Phi_{J}(y)]_{\pm,x_{0}=y_{0}} = -i \delta^{I}_{J} \delta(\boldsymbol{x} - \boldsymbol{y}), \\
        & [\Phi_{I}(x), \Phi_{J}(y)]_{\pm,x_{0}=y_{0}} = [\pi^{I}_{\Phi}(x), \pi^{J}_{\Phi}(y)]_{\pm,x_{0}=y_{0}} = 0
    \end{aligned}
\end{equation}
where $\Phi_{I} = (B^{a}, A_{k}^{a},c^{a},\bar{c}^{a},\psi^{a})$ and $\pi_{\Phi}^{I}$ denote the respective conjugate momenta, by the same order.
The new gauge fixing term \eqref{eq:gauge_fixing_term_canonical} spoils the local gauge invariance, but it is possible to identify a generalized gauge transformation that leaves $\mathcal{L}_{\text{eff}}$ unchanged, the BRST tranformation \cite{BECCHI1976287,tyutin2008gauge} such that:
\begin{equation}
\begin{aligned}
    \label{eq:brst_transform}
    & \delta A^{a}_{\mu} = (D_{\mu}c^{a}), \quad \delta c^{a} = -\frac{1}{2}f^{abc}c^{b}c^{c}, \quad \delta \bar{c}^{a} =  i\left(B^{a}\right), \\
    & \delta \psi^{a} = i\left(t^{c}\right)^{ab}c^{c}\psi^{b}, \quad \delta B^{a} = 0
\end{aligned}
\end{equation}
The transformation law for a given field $\Phi$ is:
\begin{equation}
\begin{aligned}
    \Phi \longrightarrow \Phi' = \Phi + \delta \Phi, \\
    \Phi = \{A^{a}_{\mu}, c^{a}, \bar{c}^{a}, \psi^{a}, \bar{\psi}^{a}\}.
\end{aligned}
\end{equation}
This transformation allows us to write the respective conserved Noether current $(J_{B})_{\mu}$ \cite{KugoOjima79}, which can be rewritten using the equations of motion as:
\begin{equation}
    \label{eq:conserved_current}
    (J_{B})_{\mu} = \sum_{\Phi_{I}} \delta \Phi_{I} \frac{\partial \mathcal{L}}{\partial(\partial_{\mu}\Phi_{I})} = B^{a}D_{\mu}c - (\partial_{\mu}B^{a})c^{a} + \frac{i}{2}g(\partial_{\mu}\bar{c}^{a})f^{abc}c^{b}c^{c}- \partial^{\nu}(F_{\mu \nu}^{a} c^{a})
\end{equation}
from which the respective conserved charge can be derived:
\begin{equation}
\begin{aligned}
    \label{eq:brst_charge_hermitian}
    & Q_{B} = \int d^{3}x (J_{B})_{0}(x) = \int d^{3}x \left[ B^{a} \left( D_{0}c\right)^{a} - \dot{B}^{a}c^{a} + \frac{i}{2}g\dot{\bar{c}}^{a}f^{abc}c^{b}c^{c} \right] \\ \implies 
    & Q_{B}^{\dagger} = Q_{B}.    
\end{aligned}
\end{equation}
This charge generates the BRST transformation:
\begin{equation}
    \label{eq:brst_generated}
    [iQ_{B}, \Phi]_{\pm} = \delta \Phi
\end{equation}
where the commutation (anti-commutator) should be taken if $\Phi$ has a even (odd) number of ghost fields\footnote{Due to the fermionic nature of ghost fields.}. Using this fact, a very important relation regarding $Q_{B}$ can be derived:
\begin{equation}
    \label{eq:QB Nilpotent}
    \{Q_{B},Q_{B}\}= 2 Q_{B}^2 = 0
\end{equation}
in other words, the BRST-charge is nilpotent. This nilpotency of $Q_{B}$ will be relevant for the confinement mechanism that we will see later. 
Additionally, the new hermitian assignment of the ghost fields allows the identification of a scale symmetry \cite{KugoOjima79}:
\begin{equation}
\begin{aligned}
    \label{eq:fp_symmetry}
    c^{a} \longrightarrow e^{\theta}c^{a}, \\
    \bar{c}^{a} \longrightarrow e^{-\theta}\bar{c}^{a}
\end{aligned}
\end{equation}
which has its own charge as well, the FP-charge:
\begin{equation}
    \label{eq:fp_charge}
    Q_{c} = i \int d^{3}x \left( \bar{c}^{a}\left(D_{0}c\right)^{a} - \dot{\bar{c}}^{a}c^{a}\right) = Q^{\dagger}_{c}. 
\end{equation}
Computing the commutation relations between the conserved FP-charge and the ghost fields yields:
\begin{equation}
\begin{aligned}
    \label{eq:FP Charge Commutator}
    [iQ_{c}, c^{a}(x)]_{\pm} & = c^{a}(x), \\
    [iQ_{c}, \bar{c}^{a}(x)]_{\pm} & = -\bar{c}^{a}(x)
\end{aligned}
\end{equation}
\noindent
These relations can be verified using the canonical commutation relation for the ghost fields. Additionally, this shows $iQ_{c}$ has integer eigenvalues\footnote{While this may be surprising, it is consistent with the indefinite metric of the state space $\mathcal{V}$. Let: 
\begin{equation}
\begin{aligned}
    \label{eq:imaginary_eigenvalues}
    \hat{A}^{\dagger} = \hat{A}, \quad \hat{A} \mid \alpha_{n} \rangle = a_{n} \mid \alpha_{n} \rangle \implies \\
        \langle \alpha_{m} \mid \hat{A} - \hat{A}^{\dagger} \mid \alpha_{n} \rangle = \left(a_{m}^{*} - a_{n}\right) \langle \alpha_{m} \mid \alpha_{n} \rangle = 0 
\end{aligned}
\end{equation}
For $n=m$, $\langle \alpha_{n} \mid \alpha_{n} \rangle$ does not need to be strictly positive. So, we can't draw the conclusion that $a_{n}^{*}=a_{n}$, which means in general that the eigenvalues of an Hermitian operator can be complex. Furthermore, \eqref{eq:imaginary_eigenvalues}  implies that the eigenvalues of a hermitian operator appear in pairs conjugate of one another.}. 
The eigenvalue of $iQ_{c}$ is defined as the \emph{FP-Ghost number} $N_{FP} \in \boldsymbol{Z}$.
\section{Defining the Hilbert Space}%
\label{sec:Physicality Criteria}
The quantization of gauge theories in covariant gauges requires considering an indefinite metric (i.e negative norm states) in the state vector space $\mathcal{V}$ \cite{Nakanishi}. The existence of negative norms prevents us from interpreting inner products as probability amplitudes. Only when restricting $\mathcal{V}$ to a physical subspace $\mathcal{V}_{\text{phys}} \subseteq \mathcal{V}$, a Hilbert space with positive definite metric can be defined, enabling us to recover the usual probabilistic
interpretation of Quantum Mechanics\footnote{The Gupta-Bleuler quantization \cite{Gupta1950,Bleuler1950} of QED is a known example of this.}.  To define such space, we require that \cite{KugoOjima79}:
\begin{enumerate}
    \item The Hamiltonian of the theory $H$ is hermitian: $H^{\dagger} = H$;
    \item The physical subspace $\mathcal{V}_{\text{phys}}$ is invariant under time development;
    \item The inner product in $\mathcal{V}_{\text{phys}}$ is positive semi-definite.
\end{enumerate}
Should these requirements be satisfied, a valid quantum theory, with a physical S-matrix may be formulated on the following quotient space:
\begin{equation}
    \label{eq:Physical Subspace}
    \mathcal{H}_{\text{phys}} \equiv \mathcal{V}_{\text{phys}}/\mathcal{V}_{0}
\end{equation}
where $\mathcal{V}_{0}$ is the zero-norm subspace of $\mathcal{V}_{\text{phys}}$:
\begin{equation}
    \label{eq:zero_norm_subspace}
    \mathcal{V}_{0} = \{ | \psi \rangle \in \mathcal{V}_{\text{phys}} : \langle \psi \mid \psi \rangle = 0\}
\end{equation}
Taking the quotient space as in \eqref{eq:Physical Subspace} allows for the definition of a space with a positive-definite inner product where the only zero-norm vector is the null vector. 
It is relevant to note the first and second condition may be stated in terms of the $S$-matrix:
\begin{enumerate}
    \item $S^{\dagger}S = SS^{\dagger} = 1$ (Unitary S-Matrix);
    \item $S \mathcal{V}_{\text{phys}} = S^{-1}\mathcal{V}_{\text{phys}} = \mathcal{V}_{\text{phys}}$. 
\end{enumerate}
 The anti-ghost redefinition in \eqref{eq:ghost_anti_hermitian} is vital to the first requirement, since it allows for $\mathcal{L}^{\dagger} = \mathcal{L} \implies H^{\dagger} = H$.
\subsection{Physical Subspace and BRST-Algebra}%
\label{sub:Physical Subspace and BRS Algebra}
To specify the physical subspace, Kugo and Ojima \cite{KugoOjima79} proposed the following subsidiary condition:
\begin{equation}
    \label{eq:Subsidiary Condition}
    \mathcal{V}_{\text{phys}} \equiv \operatorname{ker} Q_{B}= \left\{ | \psi \rangle \in \mathcal{V} : Q_{B} |\psi \rangle = 0 \right\}
\end{equation}
Under the assumption that $Q_{B}$ is unbroken\footnote{An unbroken generator of symmetry annihilates the vacuum, $Q_{B}| 0 \rangle = 0$, so that the generated symmetry is not spontaneously broken. An equivalent statement is that the charge $Q_{B}$ is well-defined. This is desirable so that the BRST-Symmetry remains a global symmetry of the Lagrangian.}, this definition of the physical subspace immediately satisfies the second requirement in section \ref{sec:Physicality Criteria}, as the charge is a conserved scalar quantity, making $\mathcal{V}_{\text{phys}}$ Poincaré invariant (and in particular, invariant under time translations).
Using the equal-time commutation relations \eqref{eq:commutation_relations} and the transformation law \eqref{eq:brst_generated}, we can derive the algebra between the two conserved charges we've defined so far:
\begin{equation}
    \label{eq:brs_algebra}
\begin{aligned}
    \{Q_{B},Q_{B}\} & = 2 Q_{B}^2 = 0;\\
    [iQ_{c},Q_{B}] & = Q_{B}; \\
    [Q_{c},Q_{c}] & = 0.
\end{aligned}
\end{equation}
The first equation expresses the already mentioned nilpotency of $Q_{B}$. The second equation expresses an important fact: $Q_{B}$ carries a FP-ghost number $N_{FP}$ of one.
\subsection{Inner Product Structure}%
\label{sub:Inner Product Structure}
Finally, the positivity of the physical subspace $\mathcal{V}_{\text{phys}}$, the third requirement mentioned in section \ref{sec:Physicality Criteria} can be discussed. Using the BRST-algebra some features regarding the inner product can be derived: since $Q_{c}$ has imaginary eigenvalues while being a hermitian operator\footnote{See footnote in the previous page.}, in $\mathcal{V}$ rise orthogonality conditions such as:
\begin{equation}
    \label{eq:Strange Orthogonality Relations}
    \langle k', N'_{FP} | k, N_{FP} \rangle \propto \delta_{N_{FP}, - N'_{FP}}
\end{equation}
where $k$ and $k'$ refer to other relevant quantum numbers. The states $| k, N_{FP} \rangle$ and $| k, -N_{FP} \rangle $ are referred as \emph{FP-Conjugates} of each other. 
The existence of a nilpotent operator such as \eqref{eq:QB Nilpotent} splits the whole state space $\mathcal{V}$ into the following \cite{KugoOjima79}:
\begin{enumerate}
    \item \textbf{BRST-Singlets}: States $|\psi \rangle$ belonging to $\mathcal{V}_{\text{phys}}$ (i.e $Q_{B}|\psi \rangle =0$) such that there is no state $| \phi \rangle $ in $\mathcal{V}$ satisfying $Q_{B} |\psi \rangle = |\phi \rangle$. Using the FP-Ghost number, it is possible to divide these into:
        \begin{itemize}
            \item Physical States - BRS-Singlet with $N_{FP} = 0$;  
            %\item Unpaired Singlet ($N_{FP} \neq 0$) - States whose FP-Conjugate is a member of a Doublet;   

            \item Paired Singlet ($N_{FP} \neq 0$) - States that are FP-Conjugate of each other. 
        \end{itemize}
    \item \textbf{BRST-Doublets}: Pair of states $(|\psi \rangle$, $|\phi \rangle)$ that are related by $Q_{B} |\psi \rangle = |\phi \rangle$. The state $|\phi \rangle$ is called a \emph{daughter} state whose \emph{parent} is $|\psi \rangle$. The corresponding FP-Conjugate of these pairs form another doublet, the set of these 4 states constitutes the \emph{quartet}. 
\end{enumerate}
It is possible then to draw the following conclusions:
\begin{enumerate}
\item The paired singlets are excluded since the existence of such states would spoil the norm-positivity in $\mathcal{V}_{\text{phys}}$. The BRST-algebra alone is not enough to exclude the existence of these states, which is why this requirement is theory-dependent. Fortunately, there is evidence to support their absence in gauge theories \cite{KugoOjima79}. 
\item A \emph{daughter} state $|\phi \rangle = Q_{B} |\psi \rangle$ belongs to the zero-norm subspace \eqref{eq:zero_norm_subspace}, but is also orthogonal to all other states in $\mathcal{V}_{\text{phys}}$ :
\begin{equation}
    \label{eq:daughter_state_orthogonal_to_vphys}
    \begin{aligned}
        & \langle \phi | \phi \rangle = \langle \psi | Q_{B}^{2} | \psi \rangle = 0 \\ 
        & \langle f | \phi \rangle = \left(\langle f | Q_{B} \right) | \psi \rangle = 0, \quad | f \rangle \in \mathcal{V}_{\text{phys}}. 
    \end{aligned}
\end{equation}
\end{enumerate}
\subsection{The Quartet Mechanism}%
\label{sub:The Quartet Mechanism}
The nilpotency of $Q_{B}$ has split the state vector space $\mathcal{V}$ into singlets and quartets. Kugo and Ojima \cite{KugoOjima79} have proven in a general basis the \emph{quartet mechanism}: states belonging to quartets are confined in the physical subspace $\mathcal{V}_{\text{phys}}$. Namely, given a quartet such as:
\begin{equation}
\begin{aligned}
\label{eq:Example_quartet}
& | k, N_{FP} \rangle = \chi_{k}^{\dagger}| 0 \rangle \equiv | \chi_{k} \rangle & \quad & | k -N_{FP} \rangle = -\beta_{k}^{\dagger}| 0 \rangle \equiv -|\beta_{k} \rangle \\
& | k, N_{FP}+1 \rangle = -i\gamma_{k}^{\dagger}| 0 \rangle \equiv -i |\gamma_{k} \rangle & \quad & | k , -N_{FP}-1 \rangle = -\bar{\gamma}_{k}^{\dagger} |0 \rangle \equiv - | \bar{\gamma}_{k} \rangle 
\end{aligned}
\end{equation}
where
\begin{equation}
    \begin{aligned}
    \label{eq:Example_quartet_2}
    | \gamma_{k} \rangle = -i Q_{B} | \chi_{k} \rangle, \quad | \beta_{k} \rangle = Q_{B} | \bar{\gamma}_{k} \rangle
    \end{aligned}
\end{equation}
and the corresponding quartet projector\footnote{For $n=0$, $P^{(0)}$ is defined to project the singlet components.}:
\begin{equation}
 \begin{aligned}
 \label{eq:Projector}
        P^{(n)} = \left(\frac{1}{n}\right) \sum_{k} \left\{-\beta_{k}^{\dagger} P^{(n-1)} \chi_{k} - \chi_{k}^{\dagger} P^{(n-1)} \beta_{k} - \sum_{j} \left(  \omega_{jk}\beta_{k}^{\dagger} P^{(n-1)} \beta_{j} \right)\right. + \\ +\left. i \gamma^{\dagger}_{k}P^{(n-1)}\bar{\gamma}_{k} - i \bar{\gamma}_{k}^{\dagger}P^{(n-1)}\gamma_{k}\right\}.
\end{aligned}
\end{equation}
where $\omega_{jk} = [\chi_{j},\chi_{k}]_{\pm}$, it is possible to re-write \eqref{eq:Projector} as \cite{KugoOjima79}:
\begin{equation}
\begin{aligned}
    \label{eq:Projector_New}
    & P^{(n)}=\left\{i Q_B, R^{(n)}\right\} \quad \text { for  } n \geq 1,\\
    & R^{(n)} \equiv i \frac{1}{n} \sum_k\left(\bar{\gamma}_k^{\dagger} P^{(n-1)} \chi_k+\chi_k^{\dagger} P^{(n-1)} \bar{\gamma}_k+\sum_j \omega_{jk} \beta_k^{\dagger} P^{(n-1)} \bar{\gamma}_j\right).
\end{aligned}
\end{equation}
Written in this form, for any two states $| f \rangle, | g \rangle \in \mathcal{V}_{\text{phys}}$:
\begin{equation}
    \label{eq:physstates_havenounphyisical_particles}
    \langle f | P^{(n)} | g \rangle = 0 \quad n \geq 1.
\end{equation}
So, quartet members do not contribute to the inner product defined in $\mathcal{V}_{\text{phys}}$, effectively confining these states. The recursive definition for the quartet projector in \eqref{eq:Projector} proves this for multi-particle states. So, as long the singlet states have positive norm, the quotient space:
\begin{equation}
    \label{eq:hilbert_space}
    \mathcal{H}_{\text{phys}} = \mathcal{V}_{\text{phys}}/\mathcal{V}_{0}
\end{equation}
is a valid Hilbert space to formulate our quantum theory, satisfying the third requirement\footnote{Note that, however, that this is only the case if the paired singlets mentioned in section \ref{sub:Inner Product Structure} can be excluded. Additionally, we assume that the singlets with $N_{FP}=0$, that we wish to use as representatives as physical states, have positive norm.} in section \ref{sec:Physicality Criteria}. The same result applies by linearity for field operators built with
the creation and annihilation operators given in \eqref{eq:Example_quartet}. Note however some assumptions must be made if this classification is to hold at the asymptotic level\footnote{See Appendix C of \cite{KugoOjima79} for the necessary assumptions.}.
The BRST-transformation law \eqref{eq:brst_transform} allows for the identification of the \emph{elementary} quartet. Indeed:
\begin{equation}
    \begin{aligned}
        \label{eq:brst_transform_long_polarization_and_anti_ghost}
        [iQ_{B}, A_{\mu}^{a}] | 0 \rangle = Q_{B}\left(iA_{\mu}^{a} | 0 \rangle\right) = \left(D_{\mu}c\right)^{a} | 0 \rangle, \\ 
        \{iQ_{B}, \bar{c}^{a}\} | 0 \rangle = Q_{B}\left(i \bar{c}^{a}| 0 \rangle\right) = i B^{a} | 0 \rangle .
    \end{aligned}
\end{equation}
In the case of Yang-Mills theory without spontaneous symmetry breaking, it is possible to prove \cite{KugoOjima79} that asymptotically, the longitudinal and scalar polarizations of the gauge field $A^{a}_{\mu}$ along with the ghost and anti-ghost $c^{a}, \bar{c}^{a}$ belong to the same quartet. The transverse modes are identified as physical states with positive norm. This implies that, for this theory, the positivity requirement is satisfied.  
\section{Color Confinement}%
\label{sub:Color Confinement Criterion}
The Lagrangian Density \eqref{eq:effective_lagrangian} is no longer gauge invariant under Local SU(3) transformations \eqref{eq:gauge_transform}. However, it is still invariant under the global version of the transformation. Therefore, the corresponding conserved \emph{color current} \cite{Ojima1978-1, KugoOjima79} can be derived:
\begin{equation}
    \label{eq:Color Current}
    J^{a}_{\mu} = \left( A^{\nu} \times F_{\mu \nu}\right)^{a} + j_{\mu}^{a} + \left(A_{\mu} \times B\right)^{a} - i \left(\bar{c} \times (D_{\mu}c)\right)^{a} + i \left(\partial_{\mu}\bar{c} \times c\right)^{a}
\end{equation}
Which allows to re-write the equation of motion of $A_{\mu}$ \cite{Ojima1978-1}:
\begin{equation}
    \label{eq:A_Equation_Of_Motion}
    gJ^{a}_{\mu} = \partial^{\nu} F^{a}_{\mu \nu} + \{Q_{B},D_{\mu}\bar{c}^{a}\}
\end{equation}
The corresponding charge is written as:
\begin{equation}
    \label{eq:Color_Charge}
    Q^{a} = \frac{1}{g}\left(G^{a} + N^{a}\right)
\end{equation}
where
\begin{equation}
    \label{eq:charges_of_color_charge}
    \begin{aligned}
    & G^{a} = \int d^{3}x \text{ } \mathcal{G}_{0}^{a}(x), & \qquad & \mathcal{G}_{\mu}^{a}(x) = \partial^{\nu}F_{\mu \nu}^{a}(x) \\ 
    & N^{a} = \int d^{3}x \text{ } \mathcal{N}_{0}^{a}(x), & \qquad & \mathcal{N}_{\mu}^{a} = \{Q_{B},(D_{\mu}\bar{c})^{a}(x)\}  
    \end{aligned}
\end{equation}
Color confinement takes place in $\mathcal{H}_{\text{phys}}$ if:
\begin{equation}
    \label{eq:q_a_vanishes_in_physical_subspace}
    \langle f | Q^{a} | g \rangle = 0, \quad | f \rangle, | g \rangle \in \mathcal{V}_{\text{phys}} 
\end{equation}
Note that $N^{a}$ vanishes\footnote{If the corresponding charge is well-defined, otherwise the integrations in \eqref{eq:charges_of_color_charge} would be ill-defined.} in the physical subspace $\mathcal{V}_{\text{phys}}$ defined by \eqref{eq:Physical Subspace}. We will need to state the following equivalence between different statements of the Goldstone Theorem \cite{KugoOjima79}:
Let $Q$ be a conserved charge of a global conserved current $J_{\mu}$. Then the following statements are equivalent:
\begin{enumerate}
    \label{enum:goldstone_theorem}
    \item $Q\text{ is a well-defined charge } \left(\text{i.e annihilates the vacuum } Q|0\rangle=0\right)$;
    \item $J_{\mu} \text{ has no discrete massless spectrum } \left( \text{i.e } \langle 0 | J_{\mu} | \Psi(p^2=0) \rangle= 0\right)$;
    \item $Q \text{ remains unbroken (no spontaneous symmetry breaking)}$.
\end{enumerate}
We assume the well-definedness of the color charge $Q^{a}$, so that the color symmetry is not-spontaneously broken. Let's start by looking at the charges $N^{a}$ as we will try to prove its well-definedness using the second statement above. We consider an arbitrary linear combination of the type $N_{\alpha} = \alpha^{a} N^{a}$. To verify the existence or not of a massless spectrum we search for states satisfying:
\begin{equation}
    \label{eq:massless_spectra_of_n}
    \langle 0 | \alpha^{a}\mathcal{N}_{\mu}^{a}(x) | \Psi(p^2=0) \rangle = \alpha^{a}\langle 0 | \{Q_{B},(D_{\mu}\bar{c})^{a}(x)\} | \Psi(p^2=0) \rangle \neq 0
\end{equation} 
The only non-vanishing matrix element is given by choosing $| \Psi \rangle$ to be a state created by the gauge field $A_{\mu}^{a}$, it is possible to evaluate the inner product above to \cite{KugoOjima79}:
\begin{equation}
    \begin{aligned}
    \label{eq:inner_product_result}
    \langle 0 | \alpha^{a}\mathcal{N}_{\mu}^{a}(x) | \Psi(p^2=0) \rangle = \alpha^{a}\langle 0 | \{Q_{B},(D_{\mu}\bar{c})^{a}(x)\} | \Psi(p^2=0) \rangle = \\ = \alpha^{a}\left(\delta^{ab} + u^{ab}\right)\partial_{\mu}D^{(+)}(x-y)
    \end{aligned}
\end{equation}
where $D^{(+)}$ is the positive frequency part of the massless Pauli-Jordan function:
\begin{equation}
    \label{eq:positive_part_pj_function}
    D^{(+)}(x-y)=-i \int \frac{d^{4}p}{(2\pi)^{3}}\delta(p^2)e^{-ip(x-y)}
\end{equation}
and the parameters $u^{ab}$ can be calculated with the following relation \cite{KugoOjima79}:
\begin{equation}
    \label{eq:Kugo-Ojima_Correlation_Function}
    \int d^{4}x e^{ip(x-y)}\langle 0 | T \left\{ (D_{\mu}c)^{b}(x)g \left(A_{\nu}\times\bar{c}\right)^{a}(y)\right\} | 0 \rangle = - u^{ab} \frac{p_{\mu} p_{\nu}}{p^2} + \cdots 
\end{equation} 
The dots denote terms that are not relevant to the pole structure of the equation. Regarding the charges $G^{a}$, due to the anti-symmetry of $F_{\mu \nu}^{a}$ it can be shown (Chapter 6, Lemma 6.1 of \cite{KugoOjima79}) that if the linear combination of these charges $G_{\alpha} = \alpha^{a}G^{a}$ is well-defined, then it must vanish (i.e $G_{\alpha} = 0$). 
Now, if 
\begin{equation}
    \label{eq:KO_criterion}
    u^{ab}(p^2 = 0) = -\delta^{ab}
\end{equation}
is satisfied, then the charges $N_{\alpha}$ are well-defined, but since $Q_{\alpha}$ is assumed to be well-defined, then $G_{\alpha}$ has to be as well and in turn means that it vanishes $G_{\alpha} = 0$. This leaves the following definition for the color charge $Q_{\alpha}$:
\begin{equation}
    \label{eq:color_charge_definition}
    Q_{\alpha} = \frac{1}{g}(G_{\alpha} + N_{\alpha})= \frac{1}{g} N_{\alpha}= \frac{\alpha^{a}}{g}\int d^{3}x\left\{Q_{B},  (D_{0}\bar{c})^{a}(x)\right\}.
\end{equation}
As previously stated before, if the charge $N_{\alpha}$ is well-defined, it vanishes in the physical subspace $\mathcal{V}_{\text{phys}}$ which, according to the previous equation, applies as well to the color charge $Q_{\alpha}$. Consequently, all states in the physical subspace $\mathcal{V}_{\text{phys}}$ (and consequently in $\mathcal{H}_{\text{phys}}$) are colorless, and color confinement \eqref{eq:q_a_vanishes_in_physical_subspace} takes place:
\begin{equation}
    \label{eq:q_alpha_vanishes_in_physical_subspace}
    \langle f | Q_{\alpha} | g \rangle = 0, \quad | f \rangle, | g \rangle \in \mathcal{V}_{\text{phys}} 
\end{equation}
So, the conditions for color confinement in $\mathcal{H}_{\text{phys}}$ come as:
\begin{enumerate}
    \label{enum:ko_criterion}
    \item $u^{ab}(p^2=0) = -\delta^{ab}$;
    \item the color charge $Q^{a}$ remains an unbroken generator of symmetry;
\end{enumerate}
then color confinement \eqref{eq:q_alpha_vanishes_in_physical_subspace} takes place in $\mathcal{H}_{\text{phys}}$. Since the color symmetry is assumed to remain unbroken, the second condition is automatically satisfied. The first condition can be studied using \eqref{eq:Kugo-Ojima_Correlation_Function} to obtain $u^{ab}(p^2)$. In the Landau gauge, the condition \eqref{eq:landau_gauge} translates to $p^{\mu} A_{\mu} (p) = 0$ in momentum space, which in turns
means that this function is transverse in the Lorentz space, allowing to write:
\begin{equation}
    \label{eq:KO_Function_landauspace}
    \int d^{4}x e^{ip(x-y)}\langle 0 | T \left( (D_{\mu}c)^{b}(x)g\left(A_{\nu}\times \bar{c}\right)^{a}(y)\right) | 0 \rangle \equiv \mathcal{U}^{ab}_{\mu \nu}  = (P_{T})_{\mu \nu}(p) u^{ab}(p^{2}) 
\end{equation}
where $(P_{T})_{\mu \nu}(p) = \left(\delta_{\mu \nu} - \frac{p_{\mu} p_{\nu}}{p^2}\right)$. Following equation \eqref{eq:KO_Function_landauspace}, contracting both sides with the transversal projector yields $u^{ab}$:
\begin{equation}
    \label{eq:contraction_uab}
    (P_{T})^{\mu \nu}(p) \mathcal{U}^{ab}_{\mu \nu}(p) = (N_{d} - 1) u^{ab}(p^2).
\end{equation}
Where $N_{d}$ is the dimension of the space-time considered. Finally, it is expected that $u^{ab}(p^2)$ is diagonal is colour space:
\begin{equation}
    \label{eq:KO_Diag_colourspace}
    u^{ab}(p^2) = u(p^2)\delta^{ab}
\end{equation}
From equation \eqref{eq:KO_Function_landauspace}, contracting the right-hand side of the equation with the longitudinal projector $(P_{L})^{\mu \nu}(p) = p^{\mu}p^{\nu}/p^2$ yields zero. Taking into consideration this with \eqref{eq:contraction_uab} and \eqref{eq:KO_Diag_colourspace}, the function $u(p^2)$ may be obtained taking the trace in both color and Lorentz spaces:
\begin{equation}
    \label{eq:ko_function}
    u(p^2) = \frac{1}{(N_{d}-1)(N^{2}-1)}\sum_{a, \mu} \mathcal{U}^{aa}_{\mu \mu}(p)
\end{equation}
The factor $N^{2}-1$ is the dimension of the SU(N) Lie Algebra. Since, in our case, the gauge group is SU(3), this factor comes as $N^2-1=8$. We are also considering a four-dimensional space-time $N_{d}=4$.
The confinement criterion specified before \eqref{enum:ko_criterion} translates into:
\begin{equation}
    \label{eq:ko_criterion_function}
    \lim_{p^2 \longrightarrow 0}u(p^{2}) = -1
\end{equation}

\chapter{Lattice Computation of the Kugo-Ojima function}%
\label{cha:Lattice Computation of the Kugo-Ojima function}
In this chapter, the methodology of the computation of the Kugo-Ojima function \eqref{eq:Kugo-Ojima_Correlation_Function} on the lattice is discussed. This chapter starts by presenting the details regarding the lattice formulation of Gauge Theories. As mentioned in section \ref{sub:Regularization}, this formulation provides a method to regularize non-perturbatively gauge theories. The basic formulas are presented along with the key details that allow for the simulation of gauge theories
in the computer. Using the lattice formulation of gauge theories, we present the lattice gauge-fixing procedure and discuss the use of Monte-Carlo methods to extract quantities of interest. Finally, the lattice computation of the Kugo-Ojima function is discussed, as well as some of its properties. 
\section{Lattice Discretization}
\label{sec:Lattice Discretization}
The first step is to discretize the Euclidean space-time into a hypercubic lattice:
\begin{equation}
    \label{eq:lattice_discretization}
    x_{\mu} = a n_{\mu}, n_{\mu} \in \{0,\dots,N(\mu)-1\}.
\end{equation}
The parameter $a$ is the \emph{lattice spacing} and $N(\mu)$ is the number of sites of the lattice in the $\mu$ direction\footnote{In this work, we use only symmetric lattices, which have the same number of sites in each direction.}. This directly imposes a momentum cut-off in our theory. To see this, let's start by noting that:
\begin{equation}
\begin{aligned}
    \label{eq:momentum_cutoff_proof}
    \text{exp}\{i p_{\mu} x^{\mu} \} = \text{exp}\{ i (p^{\mu} x_{\mu} + 2 \pi n_{\mu} )\} = \text{exp} \left\{i x_{\mu}\left( p^{\mu} + \frac{2\pi}{a}\right)\right\}. 
\end{aligned}
\end{equation}
Looking at the Fourier transform of an arbitrary scalar function, the previous equation implies:
\begin{equation}
    \label{eq:ft_periodic}
    F(p) = \int d^{4}x e^{-i p^{\mu} x_{\mu}}f(x) \implies F\left(p+ \frac{2\pi}{a}\right) = F(p)
\end{equation}
Therefore, momentum space integrations can be restricted to the Brilloin zone $-\frac{\pi}{a} < p_{\mu} \leq \frac{\pi}{a}$, effectively introducing a high-momentum regulator in the theory. This allows us to write the inverse Fourier transformation of a given lattice function $f(x)$ as: 
\begin{equation}
    \label{eq:inv_fourier_transform_lattice}
    f(x) = \int^{\frac{\pi}{a}}_{-\frac{\pi}{a}} \frac{d^{4}p}{(2\pi)^{4}} F(p) e^{i p^{\mu}x_{\mu}}.
\end{equation}
To perform lattice calculations on the computer, it is usual to consider a finite lattice with periodic boundary conditions. As a consequence of \eqref{eq:ft_periodic}, the momenta on the lattice are also discretized:
\begin{equation}
\begin{aligned}
    \label{eq:periodic_boundary_conditions}
    f(x^{\mu} + L(\mu)\hat{\mu}) = f(x^{\mu}) \implies \text{exp} \{i L(\mu) p_{\mu}  \hat{\mu}\} = 1 \implies p_{\mu} = \frac{\pi }{L(\mu)} n_{\mu}, \\ n_{\mu} = -\frac{N(\mu)}{2} + 1, \dots, \frac{N(\mu)}{2}
\end{aligned}
\end{equation}
where $L(\mu) = N(\mu)a $ and $\hat{\mu}$ is the unit vector in the $\mu$ direction:
\begin{equation}
    \label{eq:unit_vector_lorentz}
    \hat{\mu} \equiv e^{\mu} = \begin{pmatrix} \delta^{\mu}_{0} \\ \delta^{\mu}_{1} \\ \delta^{\mu}_{2}  \\ \delta^{\mu}_{3} \end{pmatrix}.
\end{equation}
The momentum $p_{\mu}$ is called the \emph{naive momentum}. It is often found that the naive momenta is not the most adequate to describe observables on the lattice, which leads to the definition of the \emph{improved momentum}\footnote{This alternative definition is motivated by calculating the propagator of a lattice regularized real scalar field theory, check Section 16.2 of reference \cite{aitchison2003gauge}.}:
\begin{equation}
    \label{eq:improved_momentum}
    q_{\mu} = \frac{2}{a}\text{sin}\left(p_{\mu} a\right).
\end{equation}
\section{Gauge Fields on the Lattice}
\label{sec:links}
Having the space-time discretized, it is important to know how to write relevant quantities such as the gauge fields $A^{a}_{\mu}$. In the discrete version of the theory, these fields are replaced by the \emph{link} variable:
\begin{equation}
    \label{eq:Link Variable}
    U_{\mu}(x) \equiv \operatorname{exp}\left\{ia g_{0} A_{\mu}\left(x + \frac{a}{2}\hat{\mu}\right)\right\} \in \operatorname{SU}(3)
\end{equation}
where the subscript in $g_{0}$ was placed to emphasize that it is the bare coupling constant. Note how this link variable, that now takes the role of the gauge fields, are elements of the Lie group SU(3) whereas the gauge fields were Lie algebra elements. The link has a motivation: it is the parallel transporter between adjacent points of the lattice, see Appendix \ref{app:parallel_transportes}.
Using \eqref{eq:Link Variable}, the gauge field $A^{a}_{\mu}$ can be written as a function of the links $U_{\mu}(x)$:
\begin{equation}
    \label{eq:A_from_link}
    A_\mu\left(x+\frac{a}{2} \hat{\mu} \right)=\frac{1}{2 i g_0}\left[U_\mu(x) - U_\mu^{\dagger}(x)\right]-\frac{1}{6 i g_0} \operatorname{Tr}\left[U_\mu(x)-U_\mu^{\dagger}(x)\right]+\mathcal{O}\left(a^2\right)
\end{equation}
where the second term has been included explicitly to preserve the traceless property of the gauge field $A_{\mu}$. Due to its directional nature, it is possible to define the following \cite{gattringer2009quantum}:
\begin{equation}
    \begin{aligned}
    \label{eq:directional_property_of_u}
        U_{-\mu}(x) \equiv U^{\dagger}(x-a\hat{\mu})
    \end{aligned}
\end{equation}
Before continuing, note how both in \eqref{eq:Link Variable} and \eqref{eq:A_from_link} the argument of the gauge field $A^{a}_{\mu}$ is a point in-between two adjacent lattice points\footnote{In practice, we will define a 4-dimensional array of points and put 4 links in each point, one for each direction $\hat{\mu}$. Using formula \eqref{eq:A_from_link} gives us the value of the gauge field in the middle of two adjacent lattice points.}.    
Another important detail is the link's gauge transformation law:
\begin{equation}
    \label{eq:Link Transformation Law}
    U_{\mu}^{G}(x) = G(x) U_{\mu}(x)G^{\dagger}(x + a\hat{\mu}). 
\end{equation}
The \emph{plaquette} is defined as a product of links along the shortest possible closed path:
\begin{equation}
    \label{eq:plaquette}
    \begin{aligned}
        P_{\mu \nu}(x) &= U_{\mu}(x) U_{\nu}(x + a \hat{\mu}) U_{-\mu} (x + a (\hat{\mu} + \hat{\nu})) U_{-\nu}(x + a\hat{\nu})\approx \\
        &\approx \operatorname{exp}\left\{\left(i g_{0}a^{2} F_{\mu \nu}(x)\right)\right\} + \mathcal{O}(a^{3}).
    \end{aligned}
\end{equation}
\begin{figure}[H]
\begin{center}
    \includegraphics[scale=1]{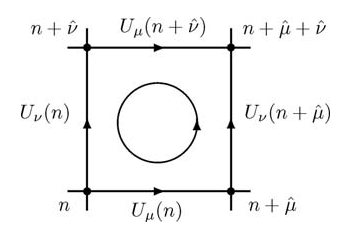}
    \caption{The Plaquette as a closed loop of links. The definition in \eqref{eq:directional_property_of_u} was used. Image from \cite{gattringer2009quantum}.}
    \label{fig:plaquette}
\end{center}
\end{figure} 
\noindent
Using \eqref{eq:Link Transformation Law}: 
\begin{equation}
    \label{eq:trace_plaquette}
    \operatorname{Tr}\left\{P_{\mu \nu}^{G}(x)\right\} = \operatorname{Tr}\left\{P_{\mu \nu}(x)\right\}
\end{equation}
due to the cyclic property of the trace and summing over all plaquettes, each counted with only one orientation, it is possible to define the \emph{Wilson Gauge Action} \cite{PhysRevD.10.2445}:
\begin{equation}
    \label{eq:Wilson Gauge Action}
    S_{G}[U] = \beta \sum_{x} \sum_{\mu < \nu} \text{Re} \left[ \text{Tr} \{ \boldsymbol{1} - P_{\mu \nu}(x)\}\right]
\end{equation}
where $\beta = \frac{2N_{c}}{g^{2}_{0}}$, and $N_{c}$ is the number of colours ($N_{c} = 3 $ in the case of QCD). In this work, we will be considering $\beta = 6.0$ which implies that the bare coupling constant is set to one (i.e $g_{0}=1$). It can be shown that definition \eqref{eq:Wilson Gauge Action} approximates to the Yang-Mills action in the continuum \cite{PhysRevD.10.2445}:
\begin{equation}
    \label{eq:Wilson Gauge Action in the Continuum}
    S_{G}[U] = a^{4}\frac{\beta}{4} \sum_{x} \sum_{\mu \nu} (F_{\mu \nu}(x))^{2} + \mathcal{O}(a^{2}).
\end{equation}
The set of links in a given lattice is called a \emph{configuration}. The definition \eqref{eq:Wilson Gauge Action} is the only action used for this work, since we shall be considering only pure-gauge configurations, since the Kugo-Ojima correlation function can be computed using these configurations. Some further details regarding the configurations are given in the next section. 
\section{Numerical Aspects}%
\label{sec:Numerical Aspects}
So far, we have discussed how to define the gluon fields on the lattice as well as how to compute the action for a given configuration. The next step, which is discussed in this section, is the calculation of quantities of interest, such as Green's functions or observables. Additionally, Green's functions depend on the gauge fixing condition so the links should also reflect the gauge fixing condition.  
\subsection{Lattice Observables}%
\label{sub:Lattice Observables}
In the continuum limit, according to the Path-Integral formalism, Green's functions can be written as:
\begin{equation}
    \label{eq:Observables in PI}
    \langle \mathcal{O} \rangle = \frac{\int \mathcal{D}\phi_{1} \dots\mathcal{D}\phi_{n} \text{  }\mathcal{O}[\phi_{1}, \dots, \phi_{n}] e^{-S_{G}[\phi_{1}, \cdots , \phi_{n}]}}{\int \mathcal{D}\phi_{1} \dots\mathcal{D}\phi_{n} \text{ }e^{-S_{G}[\phi_{1}, \cdots , \phi_{n}]}}. 
\end{equation}
On the lattice, this can be written as \cite{gattringer2009quantum}:
\begin{equation}
    \label{eq:Lattice Observables}
   \langle \mathcal{O} \rangle = \frac{\int \mathcal{D}U \text{ }\mathcal{O}[U] e^{-S_{G}[U]}}{\int \mathcal{D}U\text{ }e^{-S_{G}[U]}}.
\end{equation}
Numerically, it is very expensive to compute the exact functional integration in the equation above. Fortunately, Monte-Carlo simulations are ideal to compute multidimensional integrals such as these. This allows approximating the observable $\langle \mathcal{O} \rangle$ as:
\begin{equation}
    \label{eq:monte_carlo_integral}
    \langle\mathcal{O}\rangle_{U} = \frac{1}{N} \sum_{\{U_{n}\}'} \mathcal{O}[U_{n}].
\end{equation}
where $\{U_{n}\}'$ is a set of configurations selected in accordance with the probability distribution given by the Boltzmann factor $P[U] \propto e^{-S_{G}[U]}$. The sampling process is accomplished by a \emph{Markov chain} \cite{gattringer2009quantum}. To estimate the uncertainty of the observables computed, we use the \emph{bootstrap} method.
This method generates $N_{\text{boot}}$ samples, each composed of $N_{\text{sample}}$ randomly selected values of the observable over the configurations. For the boot sample $i$, $\mathcal{A}_{i}$ is the observable average over the selected values by that boot sample. The upper and lower limit uncertainties of a given observable are given by \cite{gattringer2009quantum}:
\begin{equation}
    \label{eq:bootstrap_uncertainties}
    \sigma_{up} = a^{*} - \langle \mathcal{A} \rangle \qquad \sigma_{down} = \langle \mathcal{A} \rangle - b^{*}
\end{equation}
where
\begin{equation}
    \label{eq:a_star_b_star}
    \frac{\#\{\mathcal{A}_{i} < a^{*}\}}{N_{boot}} = \frac{1-C}{2} \qquad \frac{\#\{\mathcal{A}_{i} < b^{*}\}}{N_{boot}} = \frac{1+C}{2}
\end{equation}
and $C$ is the confidence level. The standard value of $C$ is 0.68, which we have used in this work. In this work, we also use $N_{\text{samples}} = 10N_{\text{boot}}$.
\subsection{Gauge Fixing}%
\label{sub:Gauge Fixing} 
The restrictions imposed on the gauge fields by the gauge fixing condition \eqref{eq:gauge_fix_condition} should also be reflected on a configuration's links. 
Fixing the Landau gauge on the lattice is achieved by maximizing the following functional:
\begin{equation}
    \label{eq:functional_gauge_fix_lattice}
    F_{U}[G] = A \sum_{x,\mu} \text{Re} \left\{ \text{Tr} \left[G(x)U_{\mu}(x)G^{\dagger}(x+a\hat{\mu})  \right] \right\}
\end{equation}
where A is a normalization factor. The gauge transformed $U^{G}(x) = G(x) U(x) G^{\dagger}(x+a\hat{\mu})$ that maximizes the functional $F_{U}[G]$ is the one that satisfies \eqref{eq:gauge_fix_condition}. In other words:
\begin{equation}
    \begin{aligned}
    \label{eq:g_that_maximizes_previous}
    \frac{\delta F_{U}[G]}{\delta G^{a}(x)}\Bigg|_{G=\tilde{G}}= 0  \implies \partial^{\mu}\left(A^{\tilde{G}}\right)_{\mu}^{a}(x) = 0, \\ 
     \frac{\delta^2 F_{U}[G]}{\delta G^{a}(x)\delta^{b}G(y)}\Bigg|_{G=\tilde{G}} \approx -M[A;x,y]^{ab} \leq 0.
    \end{aligned}
\end{equation}
The proof can be found in Appendix \ref{app:gauge_fixing_functional}. Similarly to the continuum version of the theory, it is possible to define the regions $\Gamma$, $\Omega$ and $\Lambda$ but in terms of the configurations $U$:
\begin{equation}
    \begin{aligned}
        \label{eq:lattice_regions}
        & \Gamma = \{U: \partial^{\mu}A_{\mu}[U;x] = 0\}, \\ 
        & \Omega = \{U \in \Gamma: M[U] \geq 0\}, \\
        & \Lambda = \{U \in \Omega: F_{U}[\boldsymbol{I}] \geq F_{U}[G], \forall g \in \text{SU(3)}\}.
    \end{aligned}
\end{equation}
Ideally, following the discussion of section \ref{sub:Faddeevv-Popov Procedure}, one would select the global maxima of \eqref{eq:functional_gauge_fix_lattice}, more specifically, configurations belonging to $\Lambda$. This is a highly difficult global optimization task and may require the combination of local and global optimization methods. One such combination is the CEASD\footnote{Combined Evolution Algorithm Steepest Descent.} method \cite{Oliveira_2002,Oliveira_2004,silva2007gauge}, that
combines the usual Steepest Descent method \cite{daviesSD} with an Evolutionary Algorithm. Unfortunately, deploying this method for every configuration is numerically very expensive. So, in this work, we do not consider the effect of Gribov copies, utilizing only the Steepest Descent method as a local maximization method which restricts the space of configurations to $\Omega$. It should be noted that the effect of Gribov copies is small for the gluon and ghost propagators, normally within 1 to 2 standard deviations \cite{SILVA2004177,CUCCHIERI1998841}.
\section{The Lattice Kugo-Ojima function}%
\label{sec:Lattice Computation of the Kugo-Ojima correlation functiou(p^2)n}
The necessary tools of lattice gauge theories have been introduced and we can now discuss the computation of the Kugo-Ojima function \eqref{eq:KO_Function_landauspace} on the lattice.
The lattice equivalent definition of \eqref{eq:KO_Function_landauspace} is given by \cite{sternbeck2006}:
\begin{equation}
    \label{eq:KO_lat_definition}
    \begin{aligned}
        \mathcal{U}_{\mu \nu}^{a b}(p) &:= \frac{1}{V}\sum_{x, y} \sum_{c, d, e} e^{-i p \cdot(x-y)}\left\langle D_\mu^{a e} c^e(x)g_{0} f^{b c d} A_{\nu}^d(y) \bar{c}^c(y)\right\rangle_U \\
        &=\frac{1}{V}\left\langle\sum_{x, y, z} \sum_{c, d, e} e^{-i p \cdot ( x - y )} \left(D_\mu\right)^{a e}(x;z)\left(M^{-1}\right)^{e c}(z;y) f^{b c d} A_{\nu}^d(y) \right\rangle_U
    \end{aligned}
\end{equation}
where the subscript $U$ denotes the average over configurations and $g_{0} = 1$ as mentioned in Section \ref{sec:links}.
We have written explicitly all sums to be considered, dropping the repeated index convention we have been using so far. The ghost $c^e(x)$ and anti-ghost $\bar{c}^c(y)$ fields in \eqref{eq:KO_lat_definition} were replaced with the inverse of the Faddeev-Popov operator \eqref{eq:fundamental_modular_region}:
\begin{equation}
    \label{eq:ghost_propagator}
        \langle 0 | \bar{c}^{a}(x)c^{b}(y) | 0 \rangle =  \langle 0 | \left(M^{-1}\right)^{ab}(x;y) | 0 \rangle.   
\end{equation}
This correspondence can be read off directly the Faddeev-Popov term in the Lagrangian density:
\begin{equation}
    \mathcal{L}_{FP}= i\bar{c}^{a} \partial^{\mu}\left(D_{\mu}c\right)^{a} = -i\bar{c}^{a} M^{ab}c^{b}.
\end{equation}
The lattice version of the Faddeev-Popov operator can be written as \cite{Cucchieri_2018, cucchieri2018lattice}:
\begin{equation}
    \begin{aligned}
        \label{eq:lattice_fp_operator}
     M^{a b}(x;y) =\sum_\mu \operatorname{Re}\left[\operatorname{Tr}\left\{t^a, t^b\right\}\left(U_{\mu}(x)+U_{\mu}(x-\hat{\mu})  \right)\right] \delta_{x,y} - \\ -2 \sum_\mu \operatorname{Re}\left[\operatorname{Tr}\left\{t^b t^a U_{\mu}(x)\right\}\right] \delta_{x+\hat{\mu}, y}-2 \sum_\mu \operatorname{Re}\left[\operatorname{Tr}\left\{t^a t^b U_{\mu}(x-\hat{\mu})\right\}\right] \delta_{x-\hat{\mu}, y}.
    \end{aligned}
\end{equation}
Similarly, the covariant derivative in the adjoint representation has the following equivalent lattice expression \cite{zwanziger1994fundamental}:
\begin{equation}
    \label{eq:lattice_cov_derivative}
    \left(D_\mu[U]\right)^{a b}(x;y)=2 \operatorname{Re}\left[ \operatorname{Tr}\left\{t^b t^a U_{\mu}(x)\right\}\right] \delta_{x+\hat{\mu}, y}-2\operatorname{Re}\left[ \operatorname{Tr}\left\{t^a t^b U_{\mu}(x)\right\}\right]\delta_{x, y}
\end{equation}
which, up to $\mathcal{O}(a^2)$ replicates the second expression in \eqref{eq:covariant_derivatives}.
\subsection{Computational Recipe}%
\label{sub:Computational Recipe}
We start out by writing the lattice definition of $\mathcal{U}_{\mu \nu}^{ab}$ as \cite{sternbeck2006}:
\begin{equation}
    \label{eq:Sternbeck_Recipe}
    \mathcal{U}^{ab}_{\mu\nu}(p) = \left\langle \sum_xe^{-ip\cdot x}\left[D_{\mu}\psi_{b,\nu}(p)\right]^{a}(x)\right\rangle_{U} \qquad 
\end{equation}
where $\psi^{b}_{\nu}$ is the solution of the system:
\begin{equation}
    \label{eq:system_to_solve}
    [M \psi_{b,\nu}(p)]^{c}(y)  = \frac{1}{V} \left(\sum_d f^{bcd}A^{d}_{\nu}(y)\right)e^{ip \cdot y}.
\end{equation}
However, solving \eqref{eq:system_to_solve}, requires selecting a single momenta value to compute the inversion. This effectively means that the system must be solved for each individual lattice momentum, which, for large lattices, requires a lot of computational resources to compute it in a reasonable amount of time. 
So, instead of inverting the system \eqref{eq:system_to_solve} using the full plane-wave as an extended source, we use a point-source:
\begin{equation}
    \label{eq:point_source}
    [M \psi_{b,\nu}]^{c}(y)  = \left(\sum_d g_{0} f^{bcd}A^{d}_{\nu}(y)\right) \delta_{y,y_0}.
\end{equation}
where $y_{0}$ is the coordinate of the point-source. The delta function in the previous equation fixes one of the Green's function points, effectively supressing the sum in $y$. 
Using the solution of the system \eqref{eq:point_source} in \eqref{eq:Sternbeck_Recipe}, we obtain \cite{Boucaud_2005}:
\begin{equation}
    \begin{aligned}
    \label{eq:KO_lat_definition_2}
        \left\langle\sum_{x, y, z} \sum_{c, d, e} e^{-i p \cdot x} \left(D_\mu\right)^{a e}(x;z)\left(M^{-1}\right)^{e c}(z;y) \left(g_{0} f^{b c d} A_{\nu}^d(y) \delta_{y,y_0}\right)\right\rangle_U = e^{-i p \cdot y_{0}}\tilde{\mathcal{U}}^{ab}_{\mu \nu}(p).
    \end{aligned}
\end{equation}
Note, however, that the result of this procedure is a \emph{point-to-all} propagator ($\tilde{\mathcal{U}}^{ab}_{\mu \nu}$) instead of a \emph{all-to-all} propagator ($\mathcal{U}^{ab}_{\mu \nu}$):
\begin{equation}
    \begin{aligned}
        \label{eq:point_to_all_stuff}
         & \tilde{\mathcal{U}}^{ab}_{\mu \nu}(p) = e^{ip \cdot y_{0}} \sum_{x} e^{-ip\cdot x} \text{ }\tilde{\mathcal{U}}^{ab}_{\mu \nu}(x-y_{0}), \\
         & \tilde{\mathcal{U}}_{\mu \nu}^{ab}(x-y) = \sum_{z} \sum_{c,d,e} \left(D_\mu\right)^{a e}(x;z)\left(M^{-1}\right)^{e c}(z;y) g_{0} f^{b c d} A_{\nu}^d(y)
    \end{aligned}
\end{equation}
If one included all possible sources in the calculation, we would re-gain the all-to-all propagator:
\begin{equation}
    \frac{1}{V}\sum_{y}e^{ip \cdot y}\text{ } \tilde{\mathcal{U}}^{ab}_{\mu \nu}(p) = \mathcal{U}^{ab}_{\mu \nu}(p)    
\end{equation}
but doing so would be very time consuming for the lattice volumes we will be considering. So, we will use the point-to-all propagator to obtain an estimative of the all-to-all propagator.
Since not all points are used to compute the function, the statistical fluctuations are higher comparatively to solving \eqref{eq:system_to_solve}.
Including more sources and averaging over the result allows us to obtain a better estimative of the \emph{all-to-all} propagator, improving the precision of the results.
\par
\noindent
The next task is to solve the system \eqref{eq:point_source}. The Conjugate-Gradient method \cite{barrett1994templates} seems ideal to solve the system \eqref{eq:point_source}, as the matrix $M$ is real and symmetric\footnote{Note that since the source used to invert the system \eqref{eq:point_source} is real, the solution obtained by the Conjugate-Gradient method is also real.}. However, it is singular, since a constant vector is a null-mode of the matrix\footnote{For any constant vector $C^{b}_{y}$, using \eqref{eq:ghost_propagator}:\begin{equation} M^{ab}_{xy}C^{y}_{b} = 0 \end{equation} As expected, since $M$ is the discrete version of the Faddeev-Popov operator $\partial^{\mu}D_{\mu}^{ab}$.}. In other terms, the kernel of $M$ is non-trivial \cite{Suman1996}:
\begin{equation}
    \label{eq:define_m_kernel}
    \operatorname{ker} M \equiv K = \left\{\omega: \sum_{y,b}M^{ab}(x;y)\omega^{b}(y)= 0\right\} \neq \{0\}. 
\end{equation}
as constant modes belong to this subspace. Therefore, it is only possible to invert the Faddeev-Popov matrix in the kernel's orthogonal component $K^{\perp}$.
To garantee that the solution of \eqref{eq:system_to_solve} converges to a unique solution in $K^{\perp}$, we instead solve the expanded system:
\begin{equation}
    \begin{aligned}
        \label{eq:augmented_system}
        & M Y = M \phi_{b,\nu} & \quad \text{where} \quad& \phi_{b,\nu} = \sum_{d} f^{bcd}A^{d}_{\nu}(y)\delta_{y,y_0} \\
        & M \psi_{b,\nu} = Y. & 
    \end{aligned}
\end{equation}
The multiplication of $\phi_{b,\nu}$ by $M$ in the first system garantees that $Y$ belongs to the orthogonal subspace\footnote{Any given color-vector $\omega = \omega^{a}(x)t^{a}$ may be decomposed into a linear combination of null-modes $\omega^{a}_{0} \in K$ and the remaining modes $\omega^{a}_{1} \in K^{\perp}$: 
\begin{equation}
    \label{eq:modes_decomposition}
    \omega^{a} = \omega^{a}_{0} + \omega^{a}_{1} \implies M (\omega^{a}) = M( \omega^{a}_{0} + \omega^{a}_{1} ) = M \omega^{a}_{1} \neq 0
\end{equation}
So that $M \omega^{a} \in K^{\perp}$.} $K^{\perp}$. The Conjugate-Gradient method then garantees that a unique solution is found in $K^{\perp}$ \cite{Suman1996}. This is also valid for the second system, which analogously allows us to obtain a unique solution for $\psi_{b,\nu}$ in $K^{\perp}$. 
However, rounding numerical errors may destroy the orthogonality of the solution \cite{Boucaud_2005}. These errors can introduce null-modes back to the solution. To make sure this does not happen, we periodically remove the null-modes throughout the iterations of the Conjugate-Gradient method \cite{Boucaud_2005}. Consider the Fourier decomposition of a generic field $S(x)$:
\begin{equation}
    \label{eq:fourier_modes_decomposition}
    S(x) = \sum_{p}c(p) e^{ip\cdot y} = c(0) + \sum_{p \neq 0} c(p)e^{ip \cdot x}.
\end{equation}
Summing over $x$ on both sides:
\begin{equation}
    \begin{aligned}
    \label{eq:removing_constant_mode}
        \sum_{x}S(x) = \sum_{x}c(0) + \sum_{p \neq 0,x}c(p)e^{ip \cdot x} \iff \left(\sum_{x}S(x)\right) = V c(0) + \sum_{p \neq 0}c(p)\delta_{p,0} \implies \\
        \implies c(0) = \frac{1}{V}\sum_{x}S(x).
    \end{aligned}
\end{equation}
So, throughout the Conjugate-Gradient's iterations\footnote{Every 25 iterations to be exact.}, to account for the finite precision arithmetics, we subtract the null-mode to prevent the solution from escaping the orthogonal space:
\begin{equation}
    \label{eq:cleaned_of_null_modes}
    S(x)^{\perp} = S(x) - \frac{1}{V}\sum_{x}S(x).
\end{equation}
This concludes the discussion on how to solve the system in \eqref{eq:point_source}. After solving the system, we apply the lattice version of the covariant derivative \eqref{eq:lattice_cov_derivative}, take the Fourier transform and multiply by the exponential factor $e^{ik \cdot y_{0}}$ just like stated in \eqref{eq:KO_lat_definition_2}.
For a given value of $a, \mu$ this method yields $\mathcal{U}_{\mu \mu}^{a a}$ (no sum intended) for all the values of $p^2$ with two inversions of the Faddeev-Popov matrix. This amounts to $2 \times N_d \times (N_c^2-1) =2\times4\times8=64$ inversions required to compute the full trace. 
\par
\noindent
The computation may be then summarized in the following steps:
\begin{enumerate}
    \item Computation of the Kugo-Ojima source, $\phi_{b,\nu} = \sum_{d}f^{bcd}A^{d}_{\nu}(y) \delta_{y,y_{0}}$ where we use the following relation:
        \begin{equation}
            \label{eq:kugo_source_relation}
            \sum_{d} f_{bcd}A^{d}_{\nu}(y) = -\frac{1}{2}\operatorname{Tr}\left[\left\{\left(U_{-\nu}^\dagger(y) + U_{\nu}(y)\right) - \left(U_{-\nu}^\dagger(y) + U_{\nu}(y)\right)^\dagger\right\}[t^b,t^c]\right].
        \end{equation}
        See Appendix \ref{app:kugo_source} for full-derivation. Since we're only interested in the trace of $\mathcal{U}^{ab}_{\mu \nu}$, we set $b=a$ and $\mu = \nu$;
    
    \item Solving the expanded system \eqref{eq:augmented_system} for all the values of color index $a$. $y_0$ allows us to fix the location of the point-source\footnote{Note that $\psi_{a,\mu}$ is a color-vector defined over the lattice, i.e. $\psi_{a,\mu} \equiv \left(\psi_{a,\mu}\right)^e(y)$.}. We use the conjugate gradient methods with a pre-conditioner given by the operator $-\Delta^{-1} = \left(FT\right)^{-1}q^{-2}(p)\left(FT\right)$ \cite{sternbeck2006}, where $FT$ denotes
        Fourier Transform;

    \item  After obtaining $\psi_{a,\mu}$ for all color and Lorentz indices, we  compute the covariant derivative of the result $\left[D_{\mu}\psi_{a,\mu}\right]^{a}(x) = \sum_{b,z}\left(D_{\mu}\right)^{ac}(x;z)\left(\psi_{a,\mu}\right)^{c}(z)$;
    
    \item Taking the Fourier transform of the result and multiplying by the appropriate exponential factor yields the function for a single source, $\mathcal{U}_{\mu \mu}^{a a}(p) =e^{ip \cdot y_0} \sum_{x} e^{-ik \cdot x} \left[D_{\mu}\psi_{a,\mu}\right]^{a}(x)$;
    
    \item Repeat and increment the output of the previous step for all the values of $(a,\mu)$, which, after considering the correct normalization factors \eqref{eq:ko_function}, yields the function $u(p^2)$.
\end{enumerate}
Regarding software, we make use of Lattice Quantum Chromodynamics oriented libraries such as \emph{QDP++}, \emph{Chroma} \cite{qdppp,chroma} and the MPI based software \emph{PFFT} \cite{pfft}. 
\subsection{Longitudinal Test}%
\label{sub:Longitudinal Test}
An important feature of the Kugo-Ojima function is its transversality, since the Landau gauge implies that $\sum_{\mu}p_{\mu}A_{\mu}(p)=0$. One should test if this property holds in the lattice version of the function, since a non-zero longitudinal component may appear from the inversion of the Faddeev-Popov matrix. A simple test is to contract the Lorentz indices with the longitudinal projector $ \left(P_{L}\right)_{\mu\nu}(p) = \frac{p_{\mu}p_{\nu}}{p^2} $ and to see whether the result is zero:
\begin{equation}
    \label{eq:Longitudinal_test}
    \sum_{\mu \nu} \left(P_{L}\right)_{\mu\nu}(p) \sum_{a} \mathcal{U}^{aa}_{\mu \nu}(p) \equiv L(p) = 0.
\end{equation}
In this work, we perform the longitudinal test for both the naive and improved momentum:
\begin{equation}
    \begin{aligned}
        \label{eq:longitudinal_projectors}
        \left(P_L\right)_{\mu \nu}(p) = \frac{p_{\mu}p_{\nu}}{p^2} \quad \text{and} \quad \left(P_L\right)_{\mu \nu}(q) = \frac{q_{\mu}q_{\nu}}{q^2}.
    \end{aligned}
\end{equation}
\subsection{Lattice Artifacts}%
\label{sub:Lattice Artifacts}
As a correlation function, the Kugo-Ojima function should only depend on the magnitude of the momentum (justifying the argument $p^2$). In a continuum Euclidean space-time, the Kugo-Ojima function is invariant under rotations of the O(4) group, which includes rotations and inversions. On the lattice, this group is broken down into the H(4) group \cite{gupta1998introduction}, discretizing the rotations into multiples of $\frac{\pi}{2}$. In other words, for a given momentum
$p_{\mu} = (p_{x}, p_{y}, p_{z},p_{t})$, the Kugo-Ojima function should be invariant under permutations and sign inversions. 
Despite this, due to the nature of the Monte Carlo method, the computed function on the lattice will not have the same value for these equivalent momenta. So, to reduce this effect, for a given value of $p^2$, we average the value of the Kugo-Ojima function over all the equivalent momenta. This is called Z4 averaging. 
Conversely, it is possible to prove that a polynomial scalar lattice function can be written as a function of the H(4) invariants \cite{Soto2007}:
\begin{equation}
\label{eq:h4_invariants}
\begin{aligned}
    & p^{[n]} = \sum_{\mu =1}^{N_{d}} p_{\mu}^{n}, \quad n = \{2,4,6,8\}, \\
    & u(p^{2}) = \sum_{a} u^{aa}(p^{2}) \equiv u(p^{[2]}, p^{[4]},p^{[6]}, p^{[8]}).
\end{aligned}
\end{equation}
In the continuum, the orbits are labeled only by the $p^{2}$. So, assuming that we can expand the Kugo-Ojima function on the lattice as:
\begin{equation}
    \label{eq:h4_expansion}
\begin{aligned}
    u(p^{[2]},p^{[4]}, p^{[6]}, p^{[8]}) & \approx u(p^{[2]},0,0,0) + p^{[4]}\frac{\partial u}{\partial p^{[4]}}(p^{[2]},0,0,0) + \mathcal{O}(a^4) \equiv \\
            & \equiv A(p) + p^{[4]}B(p). 
\end{aligned}
\end{equation}
we can identify $A(p)$ as the continuum version of the Kugo-Ojima function, in a finite volume. This is called the H4 method, and allows for the removal of the dependence in the higher order invariants that break the O(4) invariance. 
Alternatively, one can use momenta points that minimize $p^{[4]}$, suppressing their contribution. These momenta points belong near the diagonal of the lattice. In this work we performed both a conical cut and a cylindric cut \cite{leinweber1999asymptotic}. The conical cut is performed with a half-angle of $20^{\circ}$ along the diagonal, keeping only the points
that lay inside the cone. The cylindrical cut follows the same reasoning, keeping the points that lay within a cylinder with unit radius along the diagonal. Additionally, all points with $q < 0.7 \text{ GeV} $ are represented, and do not have to pass the previous two cuts mentioned. Although most of the results presented will be using these momentum cuts, we later compare how both methods perform.

\chapter{Results and Discussion}%
\label{cha:Results}
In this chapter the results of the lattice computation of the Kugo-Ojima function are presented. The configurations used in this work were generated considering a 4-dimensional pure Yang-Mills action, with $\beta = 6.0 \implies a^{-1} = 1.943(47)$ GeV \cite{Bali_1993}. All the configurations were rotated to the Landau Gauge. Unless specified, all plots are represented as a function of the improved momentum $q$ \eqref{eq:improved_momentum}. As mentioned before, all the results are presented as bare quantities. 
We shall present results for 4 different symmetric lattices $(32^4,48^4,64^4,80^4)$. First we will present the result for the bare Kugo-Ojima function itself, which is the main goal of this work, followed by the longitudinal test mentioned in section \ref{sub:Longitudinal Test}. Then, some statistical considerations will be made regarding the inclusion of negative momenta, the H4 method and the statistical diferences between including more configurations and more sources. Finally, a
renormalization procedure is performed on the results. The following table presents the lattice setup considered: 
\begin{table}[H]
    \centering
    \begin{tabular}{@{}ccccc@{}}
        Lattice & \# Configs & \# Sources & \multicolumn{1}{l}{$\beta$} & \multicolumn{1}{l}{$a^{-1}$ (GeV)} \\ 
        $32^{4}$ & 300 & 6 & \multirow{4}{*}{6.0} & \multirow{4}{*}{1.943(47)} \\
        $48^{4}$ & 200 & 2 &  &  \\
        $64^{4}$ & 600 & 1 &  &  \\
        $80^{4}$ & 400 & 1 &  &  \\
    \end{tabular}
    \caption{Configuration setup of this work.}
    \label{tab:my-table}
\end{table}
\noindent
Unless specified, the presented results have the configurations referred in the table above. 
\newpage
\section{The Bare Kugo-Ojima function}
\label{sec:result_ko_function}
Below we present the plots for the bare Kugo-Ojima function for each lattice:
\begin{figure}[H] 
    \begin{subfigure}[b]{0.5\linewidth}
        \centering
        \includegraphics[width=\linewidth]{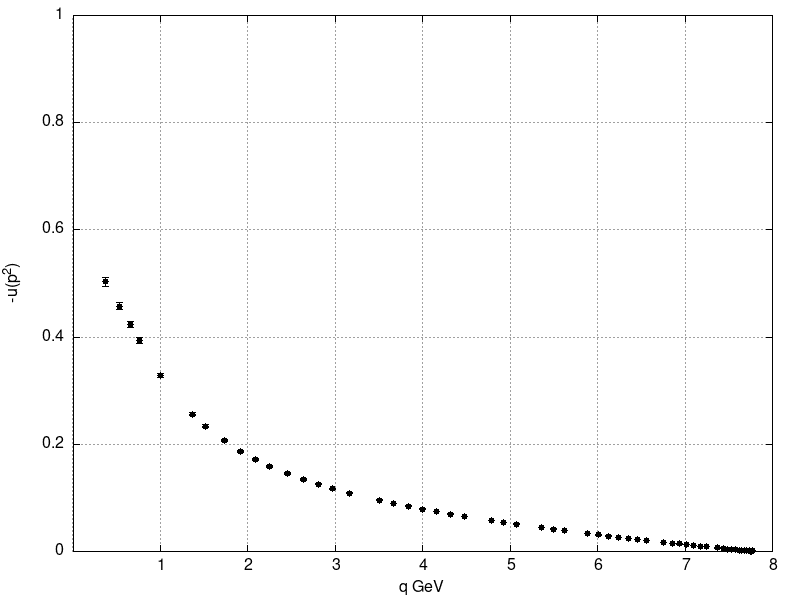} 
        \captionsetup{width=.8\linewidth}
        \caption{$32^4$ Lattice.} 
        \label{fig:ko_32}
        \vspace{4ex}
    \end{subfigure}%%
    \begin{subfigure}[b]{0.5\linewidth}
        \centering
        \includegraphics[width=\linewidth]{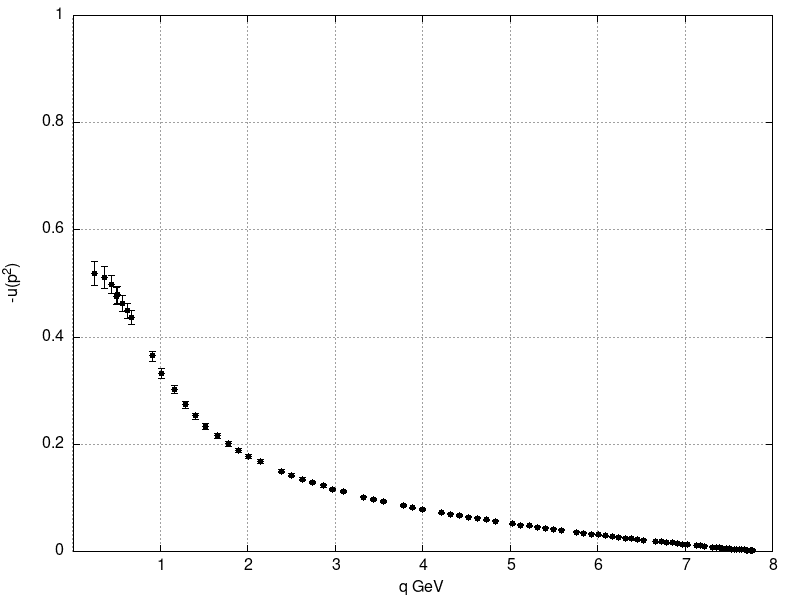} 
        \captionsetup{width=.8\linewidth}
        \caption{$48^4$ Lattice.}
        \label{fig:ko_48}
        \vspace{4ex}
    \end{subfigure} 
    \begin{subfigure}[b]{0.5\linewidth}
        \centering
        \includegraphics[width=\linewidth]{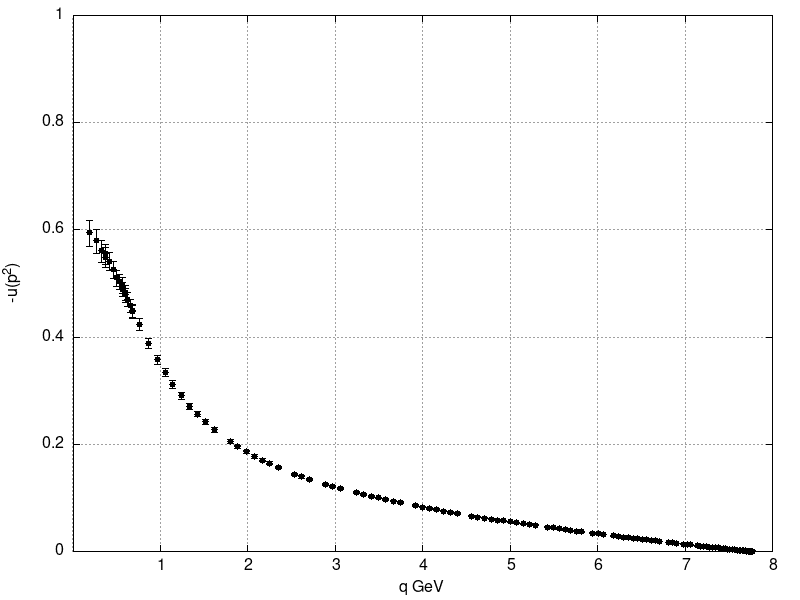} 
        \captionsetup{width=.8\linewidth}
        \caption{$64^4$ Lattice.}
        \label{fig:ko_64}
        \vspace{4ex}
    \end{subfigure}%% 
    \begin{subfigure}[b]{0.5\linewidth}
        \centering
        \includegraphics[width=\linewidth]{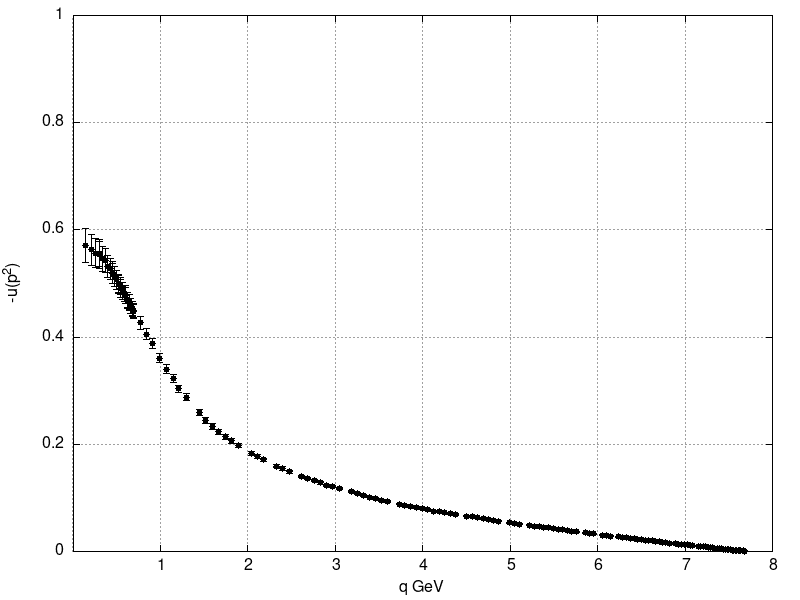} 
        \captionsetup{width=.8\linewidth}
        \caption{$80^4$ Lattice.}
        \label{fig:ko_80}
        \vspace{4ex}
    \end{subfigure} 
    \caption{The function $u(p^2)$ for all lattice volumes.}
    \label{fig:ko_all_lattices}
    \centering
\end{figure}
\newpage
\begin{figure}[H]
\begin{center}
    \includegraphics[width=0.8\linewidth]{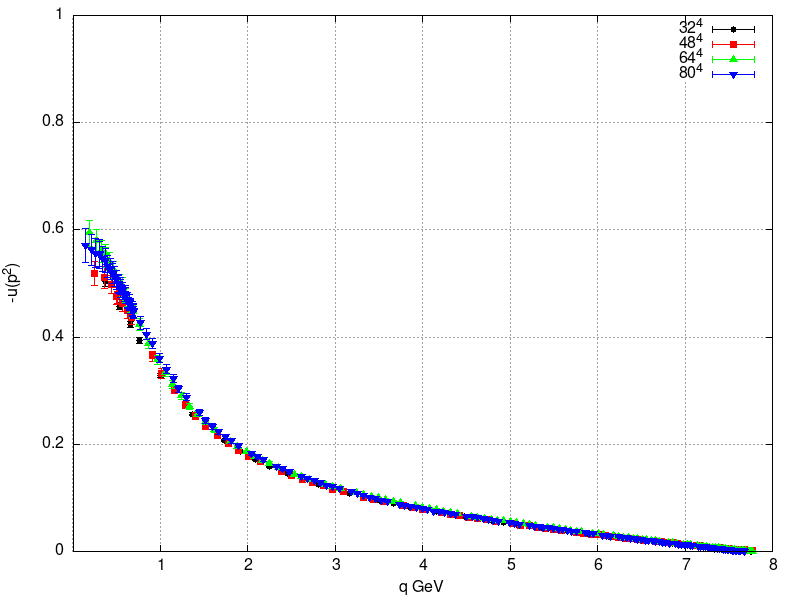}
    \caption{All the previous plots combined.}
    \label{fig:all_ko}
\end{center}
\end{figure}
\noindent
The first point for $p=0$ has been excluded from the graphs, since following the null-modes discussion \eqref{eq:fourier_modes_decomposition}, it has effectively no meaning\footnote{In \cite{Boucaud_2005} an explicit derivation for this is given, but for the non-expanded system.} \cite{Boucaud_2005}. As expected, the largest lattices have more momenta points in the infrared, making these the better choice to extrapolate the $p = 0$ value. It is helpful to zoom-in in the infrared region:
\begin{figure}[H]
    \begin{subfigure}[b]{0.5\linewidth}
    \includegraphics[width=\linewidth]{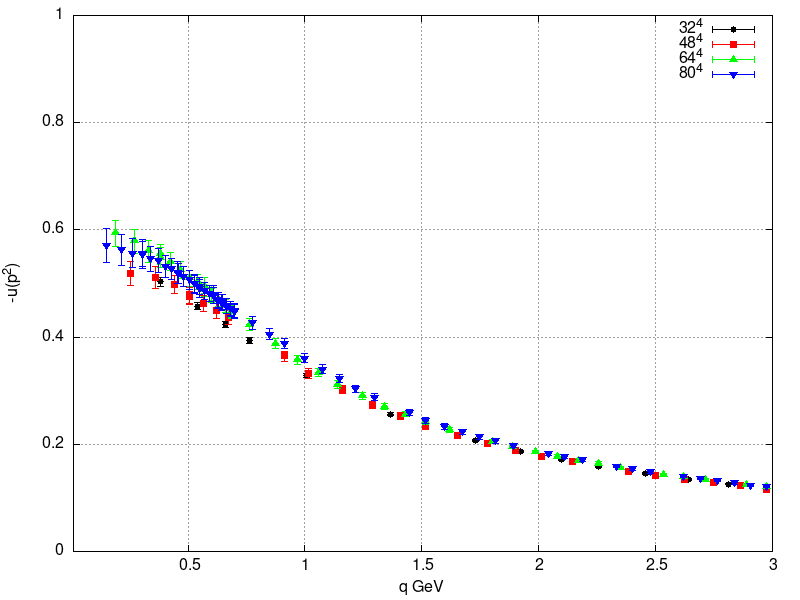}
    \end{subfigure}
    \begin{subfigure}[b]{0.5\linewidth}
    \includegraphics[width=\linewidth]{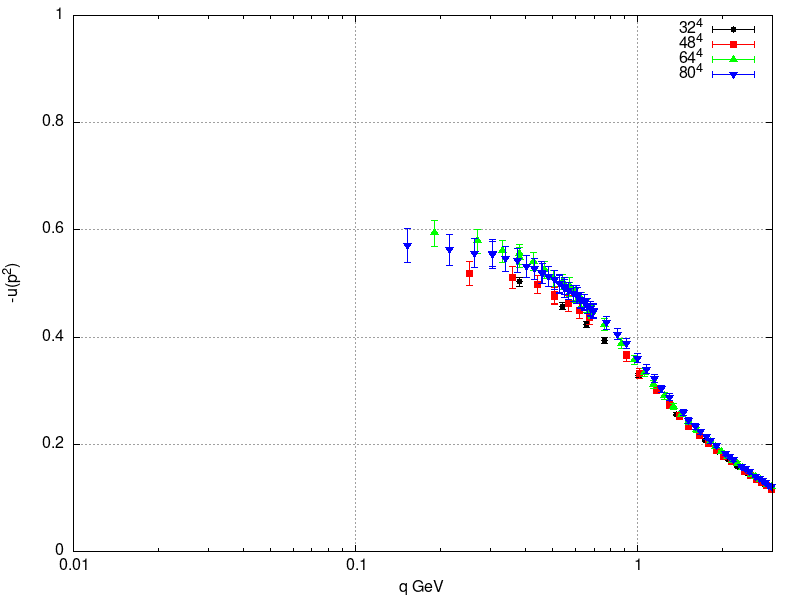}
    \end{subfigure}
    \label{fig:IR_behavior}
    \caption{Infrared behaviour of the Kugo-Ojima function.}
\end{figure}
\noindent
The volume effects appear to be small, as the data for all lattices used seem to agree within the associated uncertainties. However, it is noticeable that the fluctuations in the infrared are greater than in the ultraviolet regime, which is expected, since on the lattice there are fewer momenta points in the infrared included in the Z4 averaging process mentioned in Section \ref{sub:Lattice Artifacts}. Additionally, Gribov copies might also contribute to this effect.  
A conclusion regarding the value of $u(0)$ cannot be drawn yet before renormalizing the function, which we will consider in section \ref{sec:Renormalization of the Kugo-Ojima function}. 
\newpage
\section{Longitudinal Test}
\label{sec:longitudinal_test}
We have performed the longitudinal test mentioned in \ref{sub:Longitudinal Test} for all lattices with both types of momenta, the naive momenta and the improved momenta. In Figure \ref{fig:lq_all_lattices_2sigma} we show the results of the longitudinal tests for each lattice: 
\begin{figure}[H] 
    \begin{subfigure}[b]{0.5\linewidth}
        \centering
        \includegraphics[width=\linewidth]{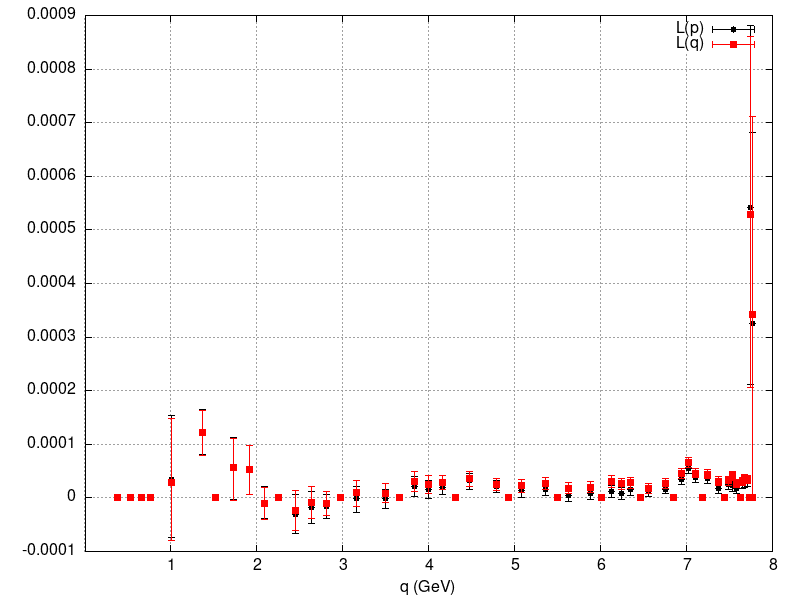} 
        \captionsetup{width=.8\linewidth}
        \caption{$32^4$ Lattice - computed with 300 configurations over 1 source.} 
        \label{fig:lt_32}
        \vspace{4ex}
    \end{subfigure}%%
    \begin{subfigure}[b]{0.5\linewidth}
        \centering
        \includegraphics[width=\linewidth]{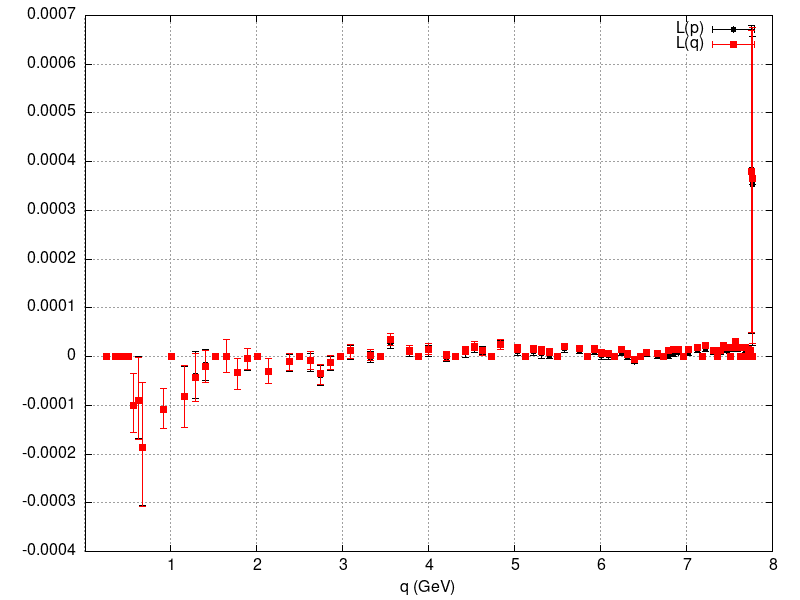} 
        \captionsetup{width=.8\linewidth}
        \caption{$48^4$ Lattice - computed with 200 configurations over 2 sources.} 
        \label{fig:lt_48}
        \vspace{4ex}
    \end{subfigure} 
    \begin{subfigure}[b]{0.5\linewidth}
        \centering
        \includegraphics[width=\linewidth]{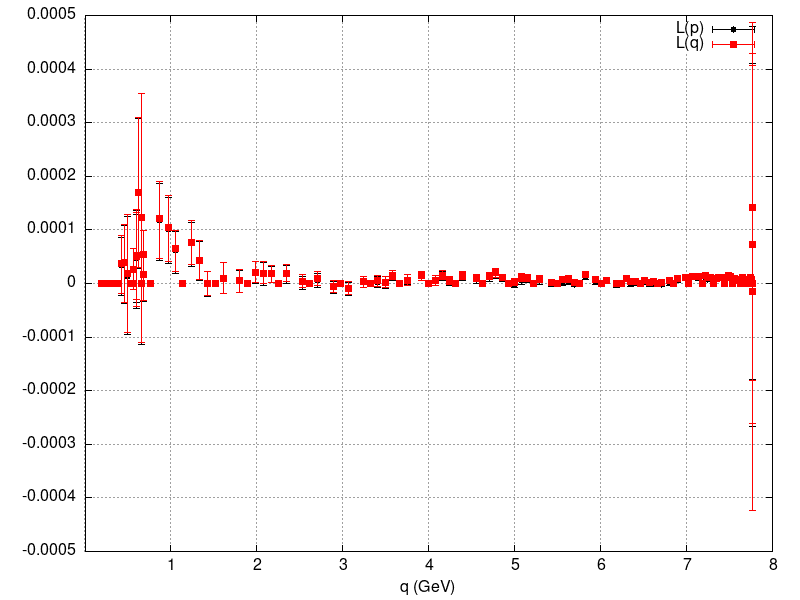} 
        \captionsetup{width=.8\linewidth}
        \caption{$64^4$ Lattice - computed with 200 configurations over 2 sources.} 
        \label{fig:lt_64}
        \vspace{4ex}
    \end{subfigure}%% 
    \begin{subfigure}[b]{0.5\linewidth}
        \centering
        \includegraphics[width=\linewidth]{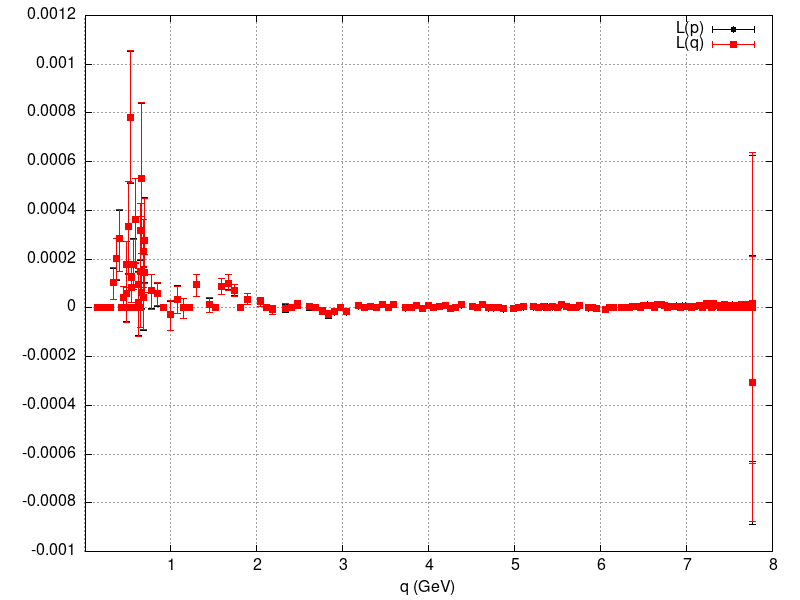} 
        \captionsetup{width=.8\linewidth}
        \caption{$80^4$ Lattice - computed with 200 configurations over 1 source.} 
        \label{fig:lt_80}
        \vspace{4ex}
    \end{subfigure} 
    \caption{Longitudinal component $L(q)$ of the lattice Kugo-Ojima function.}
    \label{fig:lq_all_lattices_2sigma}
    \centering
\end{figure}
\noindent
As we can see, the longitudinal part is negligible $(u(p^2)/L(p) \approx 0.001/0.5 =0.2\%)$ in comparison with the Kugo-Ojima function. At an uncertaintly level of $2 \sigma$, it is mostly compatible with zero, which is expected.
In the infrared, the fluctuations are significantly greater than in the ultraviolet. These fluctuations are also present in the Kugo-Ojima function itself, as mentioned in the previous section which indicates that the cause of these fluctuations is most likely correlated.
For both types of momenta, the longitudinal components are coincident for both the infrared and the ultraviolet region. In the infrared this overlap is expected since both momenta types coincide in the infrared. In the ultraviolet region, both components converge to zero\footnote{Except for the last data point in each graph. This data point corresponds to the momentum $p_{\mu} = (N(\mu),N(\mu),N(\mu),N(\mu))$. According to \eqref{eq:periodic_boundary_conditions}, this momenta is the
only of its kind when taking the Z4 average as explained in section \ref{sub:Lattice Artifacts}, which explains the large uncertainty associated.}. 
It should be mentioned however, that only the positive momenta were used in calculating the longitudinal components. 
These findings prove the transversality of the Kugo-Ojima function in the Landau gauge on the lattice. 
\newpage
\section{Inclusion of Negative Momenta Points}
\label{sec:nma_results}
Initially, the first version of the computation of the Kugo-Ojima function did not include the averages over the negative momenta points and presented a non-zero imaginary part, which is expected from \eqref{eq:KO_lat_definition_2}. The inclusion of negative momenta points in the Z4 averaging process eliminates this imaginary part and improves the statistical accuracy of the data:
%This imaginary part disappeared, with the inclusion of the negative momenta which is clear by when one considers this multiplicative factor. It also increases the statistics of the calculation, which is exactly what we wish to show with the following plots:
\begin{figure}[H] 
    \begin{subfigure}[b]{0.5\linewidth}
        \centering
        \includegraphics[width=\linewidth]{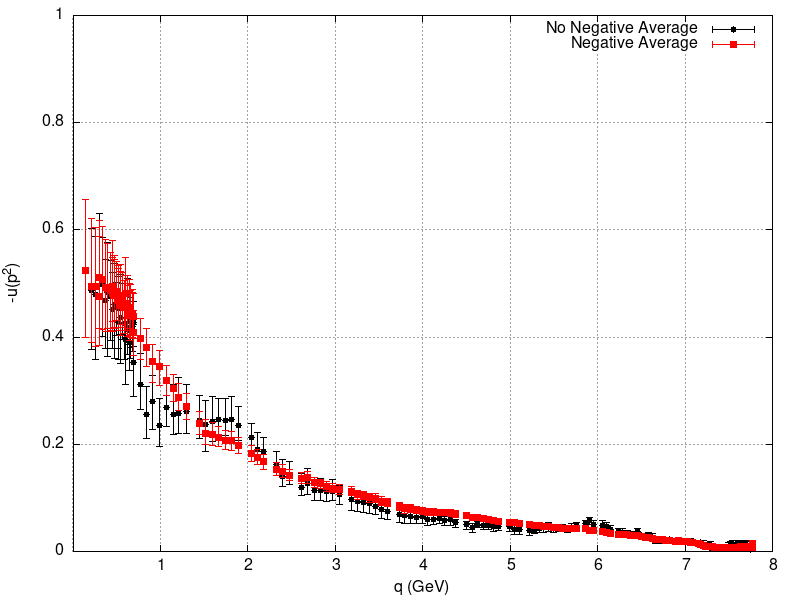} 
        \captionsetup{width=.8\linewidth}
        \caption{$80^4$ Lattice, 25 configurations}
        \label{sufig:nma_32}
        \vspace{4ex}
    \end{subfigure}%%
    \begin{subfigure}[b]{0.5\linewidth}
        \centering
        \includegraphics[width=\linewidth]{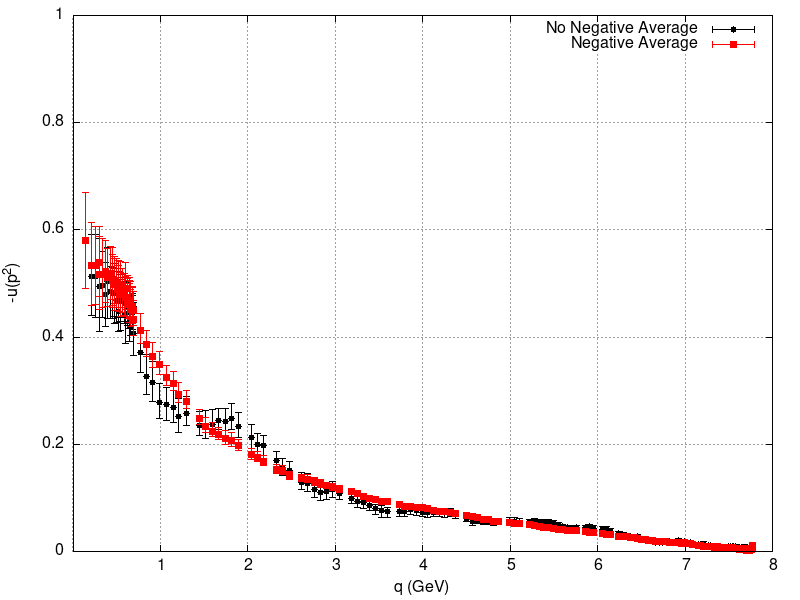} 
        \captionsetup{width=.8\linewidth}
        \caption{$80^4$ Lattice, 50 configurations}
        \label{fig:nma_48}
        \vspace{4ex}
    \end{subfigure} 
    \begin{subfigure}[b]{0.5\linewidth}
        \centering
        \includegraphics[width=\linewidth]{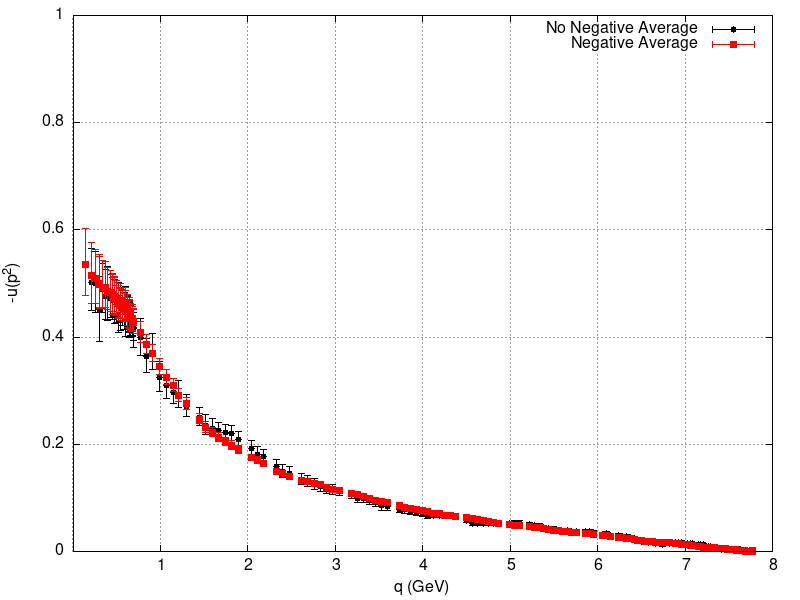} 
        \captionsetup{width=.8\linewidth}
        \caption{$80^4$ Lattice, 100 configurations}
        \label{fig:nma_64}
        \vspace{4ex}
    \end{subfigure}%% 
    \begin{subfigure}[b]{0.5\linewidth}
        \centering
        \includegraphics[width=\linewidth]{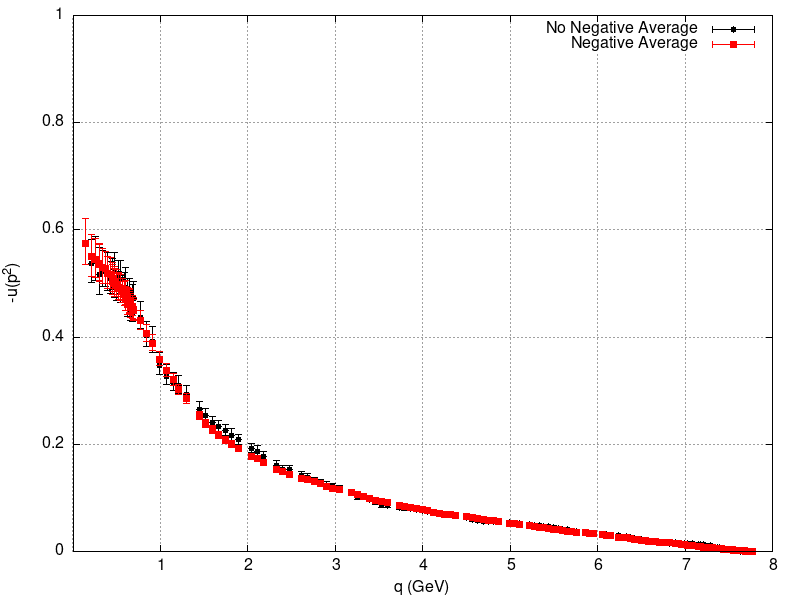} 
        \captionsetup{width=.8\linewidth}
        \caption{$80^4$ Lattice, 200 configurations}
        \label{fig:nma_80}
        \vspace{4ex}
    \end{subfigure}
    \caption{Effect of including negative momenta in Z4 averaging process for the $80^{4}$ lattice. All plots have been computed over a single source.}
    \label{fig:neg_momenta}
\centering
\end{figure}
\noindent
The difference between the two calculations decreases with the number of configurations used for the calculation, but it is noticeable the effects of including the negative momenta when inspecting the first two plots. The inclusion of negative momenta in our averaging process improves the statistical accuracy of the results, which is to be expected; if one considers only positive components for the momenta point $p_{\mu}=(p_{x},p_{y},p_{z},p_{t})$, there are only
$4!\cdot3!\cdot2!\cdot1!=24$ possible
permutations, whereas including the negative components increases this number up to $(2)^{4}\cdot4!\cdot3!\cdot2!\cdot1!=384$ momenta points included in the averaging process.   
\section{Sources vs Configurations}
\label{sec:sources_vs_configurations}
Ideally, one would compute the Kugo-Ojima function for all possible sources and take the average, which would yield the full \emph{all-to-all} propagator. However, such task is from a computational point of view, very expensive.
So, to see what kind of statistical effects additional sources have, we performed the following comparison to the $32^4$ and $64^4$ lattices:
\begin{figure}[H] 
    \begin{subfigure}[b]{0.5\linewidth}
        \centering
        \includegraphics[width=\linewidth]{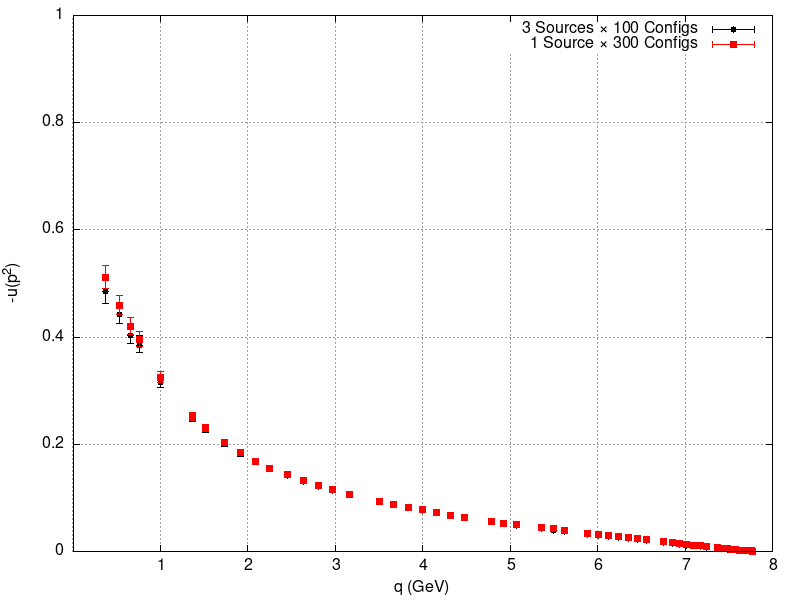} 
        \captionsetup{width=.8\linewidth}
        \caption{$u(p^2)$ computed with 300 configurations for the $32^4$ Lattice.}
        \label{fig:sources_32}
        \vspace{4ex}
    \end{subfigure}%%
    \begin{subfigure}[b]{0.5\linewidth}
        \centering
        \includegraphics[width=\linewidth]{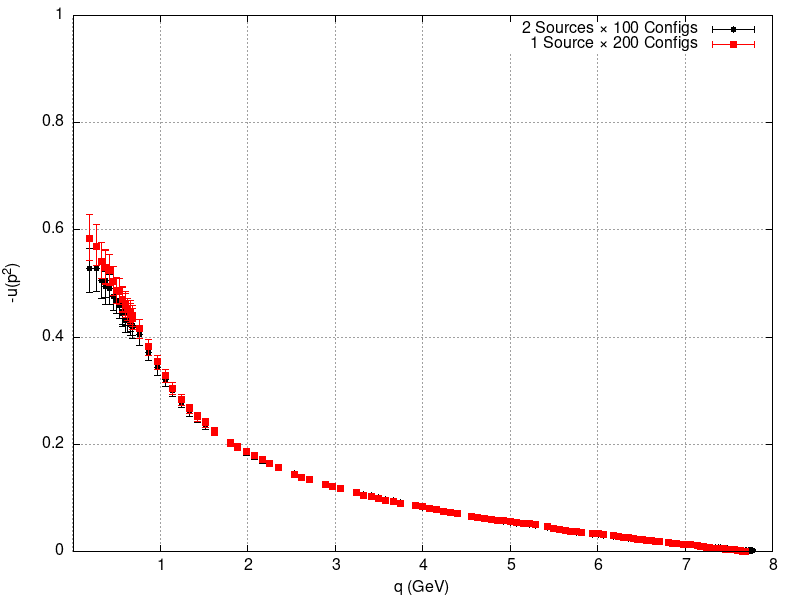} 
        \captionsetup{width=.8\linewidth}
        \caption{$u(p^2)$ computed with 200 configurations for the $64^4$ Lattice.}
        \label{fig:sources_64}
        \vspace{4ex}
    \end{subfigure} 
    \centering
    \caption{Difference between using sources vs configurations.}
    \label{fig:conf_vs_sources}
\newpage
\end{figure}
\noindent
The central values are shifted in the right panel, but the points are within one standard deviation. So, since in both cases the uncertainty intervals are similar, we can conclude that considering additional sources yield the same improvement in statistics as including more configurations in the computation. We chose to prioritize the amount of configurations over sources in our calculations, as this is more convenient from the point of view of Monte Carlos
sampling. 
\newpage
\section{H4 method}
\label{src:h4_results}
Following the discussion of section \ref{sub:Lattice Artifacts} we compare how the cuts we have been performing so far compare against the H4 method. The comparison is performed for the both the Kugo-Ojima function and its \emph{dressing function} ($K(p^{2})=-p^{2} u(p^2)$). We consider the data for the largest volume, $80^{4}$. 
\begin{figure}[H] 
    \begin{subfigure}[t]{0.5\linewidth}
        \centering
        \includegraphics[width=\linewidth]{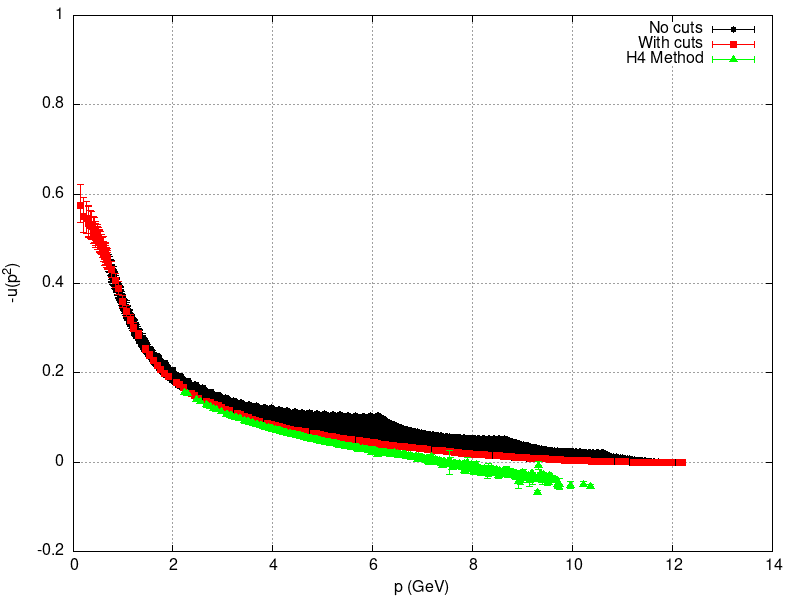} 
        \label{fig:h4_80_kop}
        \captionsetup{width=.8\linewidth}
        \caption{The Kugo-Ojima function $u(p^2)$.}
        \vspace{4ex}
    \end{subfigure}% 
    \begin{subfigure}[t]{0.5\linewidth}
        \centering
        \includegraphics[width=\linewidth]{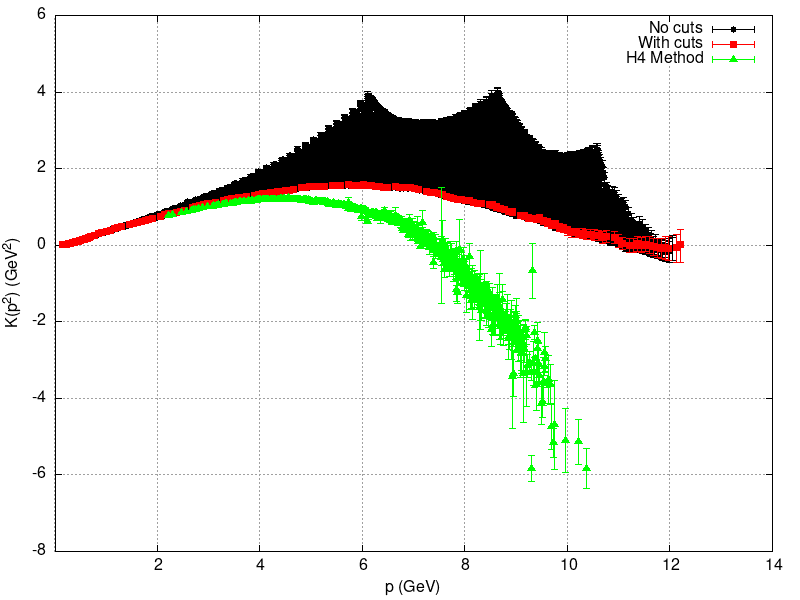} 
        \label{fig:h4_80_koq}
        \captionsetup{width=.8\linewidth}
        \caption{Kugo-Ojima's Dressing function $K(p^2)$.}
        \vspace{4ex}
    \end{subfigure} 
    \caption{Comparing the H4 method with the cuts discussed in \ref{sub:Lattice Artifacts} for the Kugo-Ojima function $u(p^2)$ and its dressing function $K(p^2)$. Both plots have been computed with 200 configurations over 1 source.} 
    \label{fig:h4_method}
\end{figure}
\noindent
As we can see, for the infrared ($q \le 3 \text{ GeV}$) the H4 method seems to agree with the canonical cuts method, but it seems to overestimate the correction of lattice artifacts outside the infrared region. Unfortunately, the method depends on finding lattice points that have the same $p^{[2]}=p^2$, which in the infrared region are fewer. 
\newpage
\section{Renormalization of the Kugo-Ojima function}%
\label{sec:Renormalization of the Kugo-Ojima function}
As mentioned in section \ref{sec:result_ko_function}, we should renormalize the bare lattice data before drawing any conclusion. To renormalize our lattice bare data we will make use of the Dyson-Schwinger equations results for $u(p^2,\mu^2)$ in \cite{Aguilar_2009,mauricio} and renormalize by setting\footnote{The results for $u(p^2,\mu^2)$ in \cite{Aguilar_2009} were renormalized using the MOM-scheme discussed in section \ref{sub:Renormalization}.}
\begin{equation}
    \label{eq:renorm_condition}
    Z_{u}(\mu^2)(1 + u_{0}(p^2)) = 1 + u(p^2,\mu^2)
\end{equation}
taking the renormalization point to be $\mu = 4.3 \text{ GeV}$. We perform the renormalization procedure for the two largest lattices $64^4$ and $80^4$. Below we present the result of the described renormalization procedure:
\begin{figure}[H] 
    \begin{subfigure}[t]{0.5\linewidth}
        \centering
        \includegraphics[width=\linewidth]{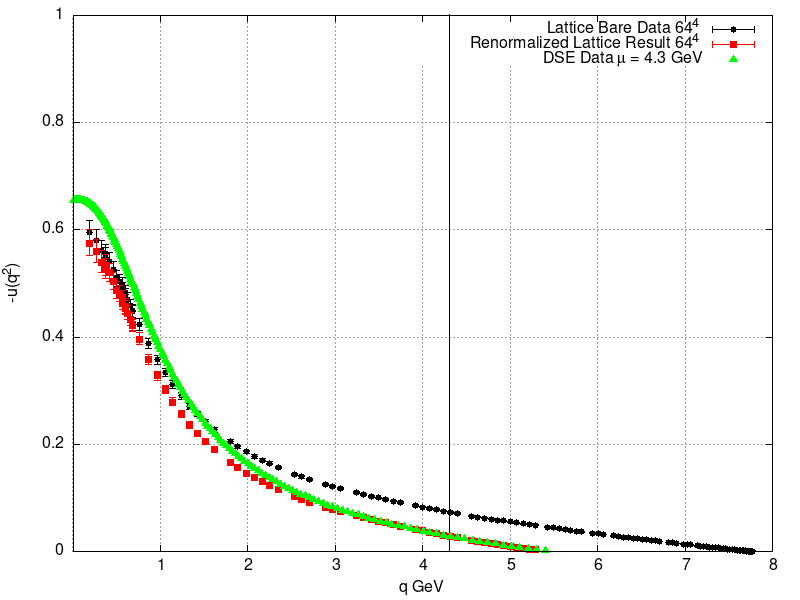} 
        \label{fig:renorm_64}
        \captionsetup{width=.8\linewidth}
        \caption{$64^4$ Lattice}
        \vspace{4ex}
    \end{subfigure}% 
    \begin{subfigure}[t]{0.5\linewidth}
        \centering
        \includegraphics[width=\linewidth]{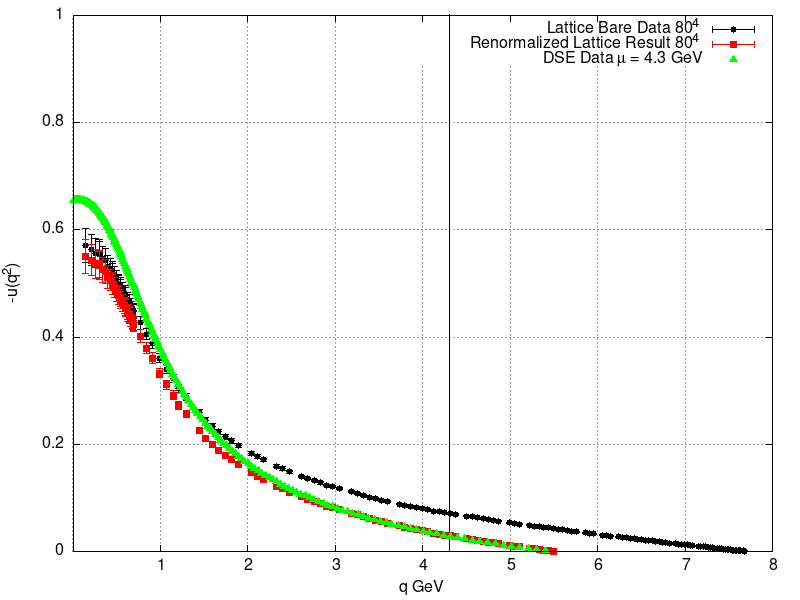} 
        \label{fig:renorm_80}
        \captionsetup{width=.8\linewidth}
        \caption{$80^4$ Lattice}
        \vspace{4ex}
    \end{subfigure} 
    \caption{The renormalized lattice Kugo-Ojima function compared with the results of \cite{Aguilar_2009}. The vertical black bar denotes the $q = 4.3 \text{ GeV}$ line.} 
    \label{fig:renorm}
\end{figure}
\noindent
It is insightful to repeat the previous plots in a logarithmic scale:
\begin{figure}[H] 
    \begin{subfigure}[t]{0.5\linewidth}
        \centering
        \includegraphics[width=\linewidth]{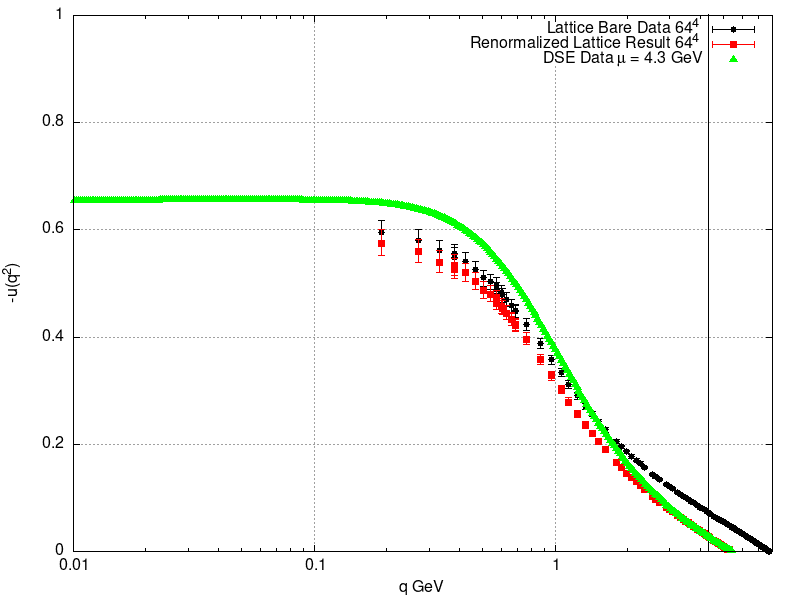} 
        \label{fig:renorm_64_log}
        \captionsetup{width=.8\linewidth}
        \caption{$64^4$ Lattice}
        \vspace{4ex}
    \end{subfigure}% 
    \begin{subfigure}[t]{0.5\linewidth}
        \centering
        \includegraphics[width=\linewidth]{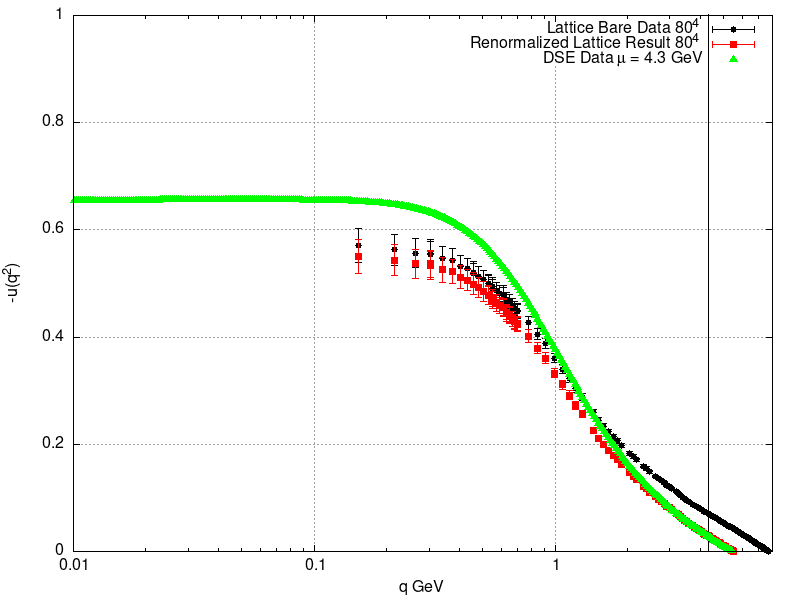} 
        \label{fig:renorm_80_log}
        \captionsetup{width=.8\linewidth}
        \caption{$80^4$ Lattice}
        \vspace{4ex}
    \end{subfigure} 
    \caption{Previous figure but in logarithmic scale.} 
    \label{fig:renorm_log}
\end{figure}
\noindent
Our results follow the same trend with the lattice data of \cite{sternbeck2006}. Our lattice results are also in qualitative agreement with the DSE result, but with a slight deviation in the infrared region. The reason for this deviation might range from approximations used in both approaches to lattice artifacts.  
\par
\noindent
The previous plots show that extrapolating to the origin would result in $u(0)\neq -1$, which points to the non-realization of the Kugo-Ojima confinement scenario. Inspecting the graphs shows that the extrapolation would result roughly in an upper bound of $u(0,\mu=4.3 \text{ GeV}) < -0.7$, which is closer to the estimate given by the author of \cite{Kondo_2009} of $u(0)=-\frac{2}{3}$. 
Furthermore, consider the following identity that relates $u(p^2)$ with the ghost dressing function \cite{Kugo1995}:
\begin{equation}
    \label{eq:ghost_and_ko_function}
    J(p^{2}) = \frac{1}{1+u(p^{2})+p^{2}v(p^2)} \qquad G^{ab}(p^{2}) = -\frac{\delta^{ab}J(p^{2})}{p^{2}}  
\end{equation}
where $v(p^{2})$ is an arbitrary function as stated in \cite{Kugo1995}. 
At $p^2 = 0$,
\begin{equation}
    \label{eq:at_0_momenta}
    J(0) = \frac{1}{1+u(0)}
\end{equation}
This relation links a diverging ghost dressing function with the realization of the Kugo-Ojima confinement scenario. This would also have to occur for the renormalized ghost dressing function as well which implies that for any renormalization point $u(0,\mu^2) = -1$, making the function RG-independent at the origin. 
The authors of \cite{binosi2009dynamics, Aguilar_2009} studied the renormalization dependence of $u(0,\mu^{2})$ and verified that there is a renormalization dependence, making it RG-dependent. 
Furthermore, there are numerical results that point to a finite ghost dressing function at the origin \cite{Boucaud_2008,Cucchieri:2008w4}, which further strenghthens that the Kugo-Ojima confinement scenario is not likely satisfied. 
All this evidence for the non-realization of the scenario could point to the non-validity of the assumption of BRST-symmetry at the non-perturbative level or at least in lattice QCD. In section \ref{sec:Physicality Criteria}, it is assumed that $Q_{B}$ remains unbroken, and it is only under that assumption that the Kugo-Ojima confinement scenario holds. This assumption is not trivial, it is not known if BRST symmetry holds at the non-perturbative regime of gauge theories and in fact, numerical evidence for BRST-symmetry
breaking on the lattice have been found \cite{cucchieri2014evidence,weishi2021non}. This invalidates the definition given in section \ref{sec:Physicality Criteria} for the physical subspace $\mathcal{V}_{\text{phys}}$, invalidating the hypothesis from the beginning.
We end by noting that the interest in the computation of the Kugo-Ojima function spans beyond just obtaining the value of $u(0)$. The function as a whole is of interest for other studies such as  the already mentioned DSE equations \cite{binosi2009dynamics,Aguilar_2009} or the Gribov-Zwazinger horizon condition \cite{zwanziger1989local,Dudal_2009,Kondo_2009}.

\chapter{Conclusion}%
\label{cha:Conclusion}
Throughout this work, we have discussed the Kugo-Ojima confinement scenario and the important role that the function $u(p^2)$ has in it. A relatively efficient method is presented for its lattice computation at all the possible momenta values. 
\par
\noindent
We start by presenting lattice bare data of the Kugo-Ojima function in the Landau gauge, studying the lattice volume dependence, the longitudinal part, which is compatible with zero, as expected. A statistical analysis of the computation, followed by the renormalized result was also presented.
\par
\noindent
Our results are in good agreement with the lattice data of \cite{sternbeck2006} and also supports the estimated provided in \cite{Kondo_2009}. The results present evidence for the non-realization of the Kugo-Ojima confinement scenario, which agrees with literature studies performed using different formalisms. These studies include the RG-dependence of $u(0,\mu^2)$ \cite{Aguilar_2009,binosi2009dynamics}, the non-enhanced ghost dressing function \cite{Boucaud_2008,Cucchieri:2008w4} and BRST-Symmetry breaking \cite{cucchieri2014evidence,weishi2021non}. This last one could be the core reason for
the non-realization of the confinement scenario. A possible reason for the non-realization of the confinement scenario is that the assumption that the BRST-charge remains unbroken is not valid. Indeed, this is not a trivial assumption, since the BRST-symmetry lacks proof of its validity at the non-perturbative level. A broken BRST-charge implies that we cannot use it to define the physical subspace as we did in Section \ref{sub:Physical Subspace and BRS Algebra}. 
\par
\noindent
In the future, efforts will be
made towards improving the statistics of the data presented and further optimization improvements can be considered. It would also be interesting to see how the Kugo-Ojima function would look on larger lattices or different gauges, although the computational resources required can be significantly high to consider other gauges.

% REFERENCES
% Edit the references.bib file to add your own references, that you can then
% \cite on your text.
\bibliographystyle{ieeetr}
\bibliography{bibliography/references.bib}
\titleformat{\chapter}[display]	% Return chapter titles to normal, taking up a whole page (cool for appendices)
{\normalfont\huge\bfseries}{\chaptertitlename\ \thechapter}{20pt}{\Huge}
\begin{appendix}			% Start appendices
\chapter{Parallel Transporter}	% One chapter per appendix
\label{app:parallel_transportes}
In this appendix, we motivate the definition of the link variable provided in \eqref{eq:Link Variable}.
Having our space discretized into a lattice, the derivative should be replaced by one of the following finite diference formula:
\begin{equation}
    \label{eq:lattice_derivative}
    \begin{aligned}
    \partial_{\mu}f(x) = \frac{f(x+a\hat{\mu}) - f(x)}{a} + \mathcal{O}(a^2), \\
        \partial_{\mu}f(x) = \frac{f(x+a\hat{\mu}) - f(x-a\hat{\mu})}{2a} + \mathcal{O}(a^3)
    \end{aligned}
\end{equation}
where $\hat{\mu}$ is the unit vector in the $\mu$ direction.
The first option should be closer to its continuum counter-part than the second, but at the expense of a less rigorous aproximation. In this work, the lattice spacing $a$ should be small enough to use the first option.
Consider an arbitrary field $\phi(x)$ that transforms under some representation of a given Lie Group G:
\begin{equation}
    \label{eq:generic_gauge_transform}
    \phi(x) \longrightarrow \phi'(x)  = G(x) \phi(x). 
\end{equation}
The corresponding kinetic term $(\partial_{\mu}\phi(x))^{\dagger}(\partial^{\mu}\phi(x))$ will have products like $\phi(x+a\hat{\\mu})^{\dagger}\phi(x)$ which, in general, explicitly violates gauge invariance. Gauge invariance is restored by introducing the parallel transporter $\Lambda(x,y)$ \cite{peskin2018}:
\begin{equation} 
    \label{eq:parallel_transporter}
    \Lambda(x,y) = P \operatorname{exp} \left(i g\int_{\mathcal{C}_{xy}} A^{\mu} ds_{\mu}\right)
\end{equation}
where the $P$ stands for \emph{path-ordered} integral\footnote{Let $s \in [0,1]$ be a parametrization of the path $\mathcal{C}_{xy}$. The path-ordering prescription will order the $A^{\mu}(x(s))$ in the exponential so that higher values of $s$ are on the left. It is similar to \emph{time-ordering} integral in the canonical treatment of field theories.}. The $g$ is the coupling constant associated with the gauge fields. The respective transformation law is given by
\cite{peskin2018}:
\begin{equation}
    \label{eq:parallel_transporter_transformation_law}
    \Lambda(x,y) \longrightarrow G(x) \Lambda(x,y) G^{-1}(y) = G(x) \Lambda(x,y) G^{\dagger}(y).
\end{equation}
Now, terms like $\phi(x)^{\dagger} \Lambda(x,y) \phi(y)$ are gauge invariant. 
For small lattice spacing:
\begin{equation}
    \label{eq:link_proof}
    \Lambda(x,x+a\hat{\mu}) = p\operatorname{exp} \left(ig \int_{x}^{x + a \hat{\mu}}A_{\mu}^{a}(x)t^{a}dx^{\mu}\right) \approx \operatorname{exp}\left(ig\frac{a}{2}A_{\mu}^{a}\left(x + \frac{a}{2}\hat{\mu}\right)t^{a}\right)
\end{equation}
which is the expression in \eqref{eq:Link Variable}. 

\chapter{Gauge Fixing Functional}
\label{app:gauge_fixing_functional}
In this appendix, we prove that optimizing the functional:
\begin{equation}
    \label{eq:functional_gauge_fix_lattice_app}
    F_{U}[G] = A \sum_{x,\mu} \operatorname{Re} \left[ \operatorname{Tr} \left\{G(x)U_{\mu}(x)G^{\dagger}(x+a\hat{\mu})\right\}\right]
\end{equation}
corresponds to fixing the Landau gauge for the input configuration.
Near the maximum of $F_{U}[G]$, we can expand the functional as:
\begin{equation}
    \begin{aligned}
    \label{eq:expand_near_max}
        F_{U}[\boldsymbol{1} + i \alpha^{a}(x)t^{a}] \approx F_{U}[1] + \frac{A}{4} \sum_{x, \mu} i \alpha^{a}(x) \operatorname{Tr}\left\{t^{a} \left[(U_{\mu}(x) - U_{\mu}(x-\hat{\mu})\right. \right. & \\
        \left. \left. - (U_{\mu}^{\dagger}(x) - U_{\mu}^{\dagger}(x- \hat{\mu})) \right]\right\}. &
    \end{aligned}
\end{equation}
A stationary point of the functional, such as a maximum, will obey:
\begin{equation}
    \label{eq:Derivative_near_max}
    \begin{aligned}    
        \frac{\partial F}{\partial \alpha^{a}(x)} = \frac{iA}{4}\sum_{\mu}\operatorname{Tr}\left\{ t^{a} \left[(U_{\mu}(x) - U_{\mu}(x-\hat{\mu}))\right.\right. & \\
        - \left.\left. (U_{\mu}^{\dagger}(x) - U_{\mu}^{\dagger}(x- \hat{\mu})) \right]\right\} & = 0.
    \end{aligned}
\end{equation}
Using the definition of link \eqref{eq:Link Variable}:
\begin{equation}
    \label{eq:using_link_definition}
    \sum_{\mu} \operatorname{Tr}\left\{t^{a}\left[A_{\mu}\left(x + \frac{a}{2}\hat{\mu}\right) - A_{\mu}\left(x-\frac{a}{2}\hat{\mu}\right)\right]\right\} + \mathcal{O}(a^{2}) = 0.
\end{equation}
Performing a Taylor expansion of the gluon fields yields:
\begin{equation}
    \label{eq:lat_landau_condition}
    \sum_{\mu}\partial_{\mu}A^{a}_{\mu}(x) + \mathcal{O}(a^2) = 0
\end{equation}
which concludes the proof. Furthermore, note that maximizing the functional $F_{U}[G]$ also means setting its second derivative to be negative. Looking at equation $\eqref{eq:lat_landau_condition}$, it is possible to conclude that the second variation of the functional $F_{U}[G]$ is the symmetry Faddeev-Popov matrix (i.e $-M[A;x,y]^{ab}$ in \eqref{eq:gribov_region}). So, maximizing this functional means restricting the space of configurations to $\Omega$.

\chapter{Point Source Generation}
\label{app:kugo_source}
In this appendix we prove the relation provided in equation \eqref{eq:kugo_source_relation}:
\begin{equation}
            \label{eq:kugo_source_relation_app}
            \sum_{c} f_{abc}A^{c}_{\mu}(x) = -\frac{1}{2}\operatorname{Tr}\left[\left\{\left(U_{-\mu}^\dagger(x) + U_{\mu}(x)\right) - \left(U_{-\mu}^\dagger(x) + U_{\mu}(x)\right)^\dagger\right\}[t^a,t^b]\right].
        \end{equation}
First, we multiply the gauge field $A_{\mu}\left(x + \frac{\hat{\mu}}{2}\right)$ by the Lie algebra commutator:
\begin{equation}
    \begin{aligned}
    A_\mu \left(x+\frac{\hat{\mu}}{2}\right)\left[t^a,t^b\right] = A_\mu \left(x+\frac{\hat{\mu}}{2}\right) i\sum_{c} f^{abc}t^{c}.
    \end{aligned}
\end{equation}
Using expression \eqref{eq:A_from_link} on the left-hand side yields:
\begin{equation}
    \begin{aligned}
        \label{eq:c_3}
        \left(\frac{1}{2ig_{0}}\left(U_{\mu}(x)-U_{\mu}^\dagger(x)\right) - \frac{\boldsymbol{1}}{6ig_{0}}\operatorname{Tr}\left\{U_{\mu}(x)-U_{\mu}^\dagger(x)\right\}\right)\left[t^a,t^b\right] = \\ = \sum_{c,d}if_{abc}A^{d}_\mu\left(x+\frac{\hat{\mu}}{2}\right)t^dt^c.
    \end{aligned}
\end{equation}
Taking the trace on both sides and using \eqref{eq:lie_algebra_su3}:
\begin{equation}
    \begin{aligned}
        \frac{1}{ig_{0}}\operatorname{Tr}\left\{\left(U_{\mu}(x)-U_{\mu}^\dagger(x)\right)\left[t^a,t^b\right]\right\} = \sum_{c} if_{abc}A^{c}_{\mu}\left(x+\frac{\hat{\mu}}{2}\right). 
    \end{aligned}
\end{equation}
The second term on the left hand side in \eqref{eq:c_3} is traceless because of the SU(3) Lie algebra \eqref{eq:lie_algebra_su3}. Performing a Taylor expansion on the right-hand side up to $\mathcal{O}(a^2)$ and multiplying by $-i$ allows us to write:
\begin{equation}
    \label{eq:taylor_expand}
    -\frac{1}{g_{0}}\operatorname{Tr}\left\{\left(U_{\mu}(x)-U_{\mu}^\dagger(x)\right)\left[t^a,t^b\right]\right\} = \sum_{c} f_{abc}\left(A^c_\mu(x) + \frac{a}{2}\partial_\mu A^c_\mu(x) + \mathcal{O}(a^2)\right).
\end{equation}
Using the same formula for $x-\frac{\hat{\mu}}{2}$, we get:
\begin{equation}
    \begin{aligned}
        -\frac{1}{g_{0}}\operatorname{Tr}\left\{\left(U_{\mu}(x - a \hat{\mu})-U_{\mu}^\dagger(x - a \hat{\mu})\right) \left[t^{a},t^{b} \right] \right\} = \\ = &
    \sum_{c} f_{abc}\left(A^c_\mu(x) - \frac{a}{2}\partial_\mu A^c_\mu(x) + \mathcal{O}(a^2)\right)
    \end{aligned}
\end{equation}
Using \eqref{eq:directional_property_of_u} and summing the last 2 equations, yields:
\begin{equation}
    \begin{aligned}
        \sum_{c} f_{abc}A^c_\mu(x) = -\frac{1}{2g_{0}}\operatorname{Tr}\left[\left\{\left(U_{-\mu}^\dagger(x) + U_{\mu}(x)\right) - \left(U_{-\mu}^\dagger(x) + U_{\mu}(x)\right)^\dagger\right\}\left[t^a,t^b\right]\right]
    \end{aligned}
\end{equation}
which concludes the demonstration. To end, we note that we will set $\beta=6.0$ which implies for $N_{c}=3$ that the bare coupling constant is set to one $g_{0}=1$.

%\chapter{Commutation relations of BRS-Singlet and BRS-Quartet states}
%\label{app:quartet_demonstrations}
%\input{appendix/quartet_demonstrations.tex}

%\includepdf[pages={-}]{path/to/appendix.pdf}
% or
%\input{appendix_file}
\end{appendix}

\end{document}